\documentclass[reprint,showkeys,amsmath,amssymb, aps, physrev]{revtex4-2}

\usepackage{graphicx}% Include figure files

\usepackage{xcolor}%
\definecolor{RED}{rgb}{1,0,0}
\definecolor{darkgreen}{rgb}{0,0.5,0}

\begin{document}

%\preprint{APS/123-QED}

\title{\textbf{Return point memory in knitted fabrics}}

\author{Elizabeth J. Dresselhaus\textsuperscript{$\dagger$}}
\email[Correspondence: ]{ej\_dresselhaus@berkeley.edu,\\ kranthi@berkeley.edu, s\_g@berkeley.edu}
\affiliation{University of California, Berkeley, CA 94720, USA}
 
\author{Sonja Hellebrand}
\thanks{These authors contributed equally to this work.}
\affiliation{University of Duisburg-Essen, Essen, D-45141 Germany}

\author{Rajyasri Roy}
\affiliation{Johns Hopkins University, Baltimore, MD 21218, USA}

\author{Kranthi K. Mandadapu\textsuperscript{*}}
\affiliation{University of California, Berkeley, CA 94720, USA}
\affiliation{Lawrence Berkeley National Laboratory, Berkeley, CA 94720, USA}

\author{Sanjay Govindjee\textsuperscript{*}}
\affiliation{ University of California, Berkeley, CA 94720, USA}

\date{\today}

\begin{abstract}
The tunable mechanical response of knitted fabrics underpins applications ranging from soft robotics and artificial muscles to morphing electromagnetic field sensors. Elasticity in fabrics emerges from the bending of yarn in the knitted structure; however, properties beyond elasticity are relatively unexplored.
Here, we demonstrate that knitted fabrics subjected to cyclic uniaxial stress exhibit significant hysteresis and the remarkable ability to ``remember" their response to previous deformations -- reminiscent of classical return point memory in magnetic systems. 
The hysteretic behavior deviates from the two standard models of hysteresis that usually apply to solid-state materials, viscoelasticity and plasticity. Thus, we develop a phenomenological extension of the Preisach model of hysteresis which well replicates our data, and discuss implications of these results on the underlying mechanisms of memory  in knitted fabrics.
\end{abstract}


\maketitle

\section{Introduction}
Knitting is a well-established method that transforms yarn, a bendable but axially stiff material, into fabric, a material with emergent stretch and bend in any direction. The knitting process also endows these materials with a tunable topology \cite{MarkandeMatsumoto2020}, which dramatically influences both the fabric shape and mechanical behavior \cite{niu2025geometric,Singal2023}. This tunability along with ease of manufacturing lend knitted fabrics  a wide variety of applications, including soft robotics \cite{sanchez2023_3Dknitting, duPasquier2024_Haptiknit}, the possibility of knitted artificial muscles \cite{maziz2017}, biomedical resorbable implants for tissue regeneration~\cite{ahmed:23}, knitted carrier materials for sensors~\cite{seyedin:15,si:23} and methods for waste-free clothing manufacturing~\cite{gojic:23}. Recent advances in knitting technology -- including knitting custom three-dimensional shapes \cite{sanchez2023_3Dknitting,guimbretiere.eq:25} and the use of smart or biodegradable fibers \cite{deugirmenci:23} -- have further expanded the potential of knitted structures, making them a focus of interdisciplinary research in solid-state physics, materials science, engineering, and design \cite{mishra:22}.

\begin{figure*}[t]
	\centering
	\unitlength=1mm
	\begin{picture}(140,125)
    \put(5,119){\includegraphics[angle = 270, width=6cm]{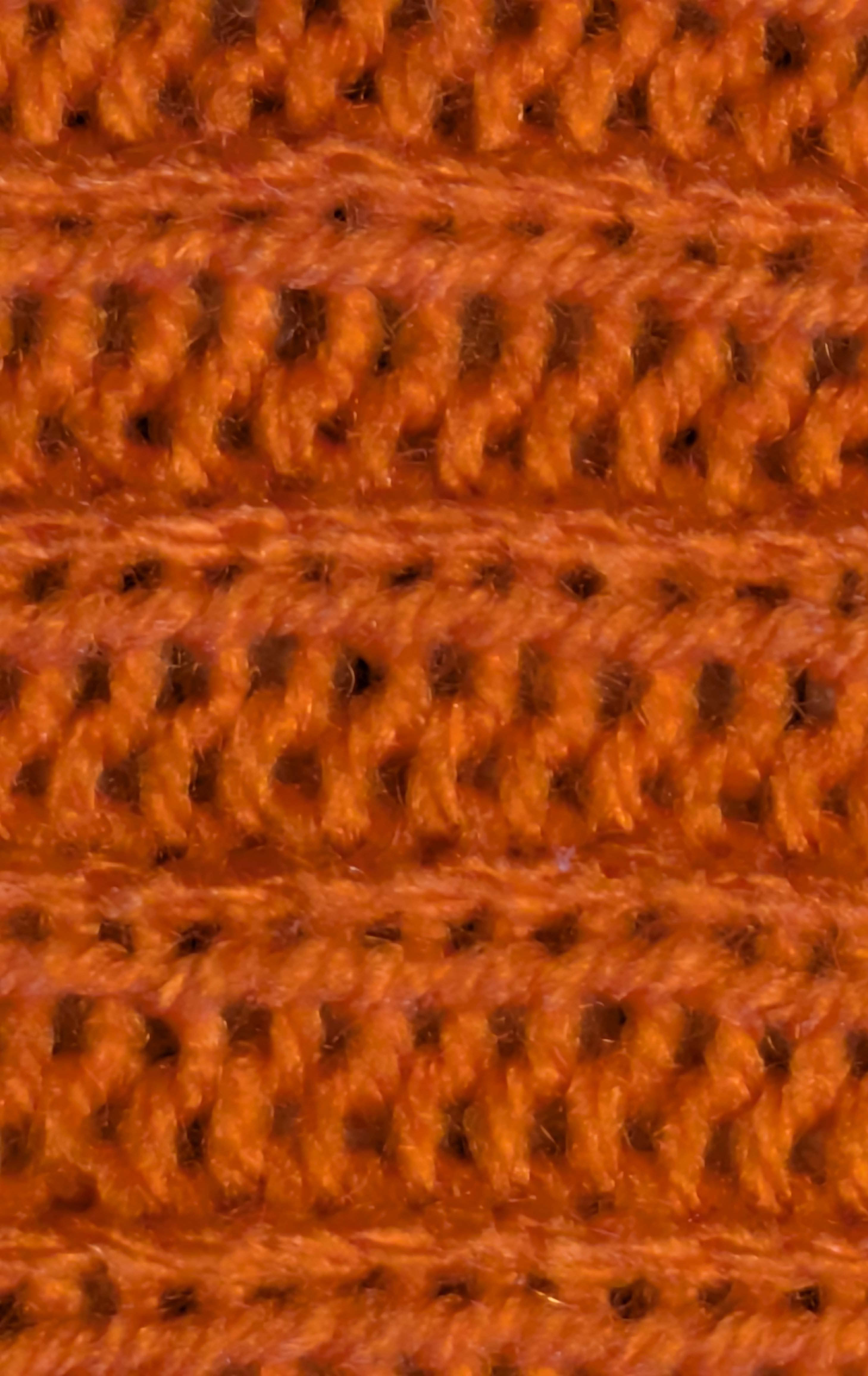}}
    \put(75,80){\includegraphics[width=6cm]{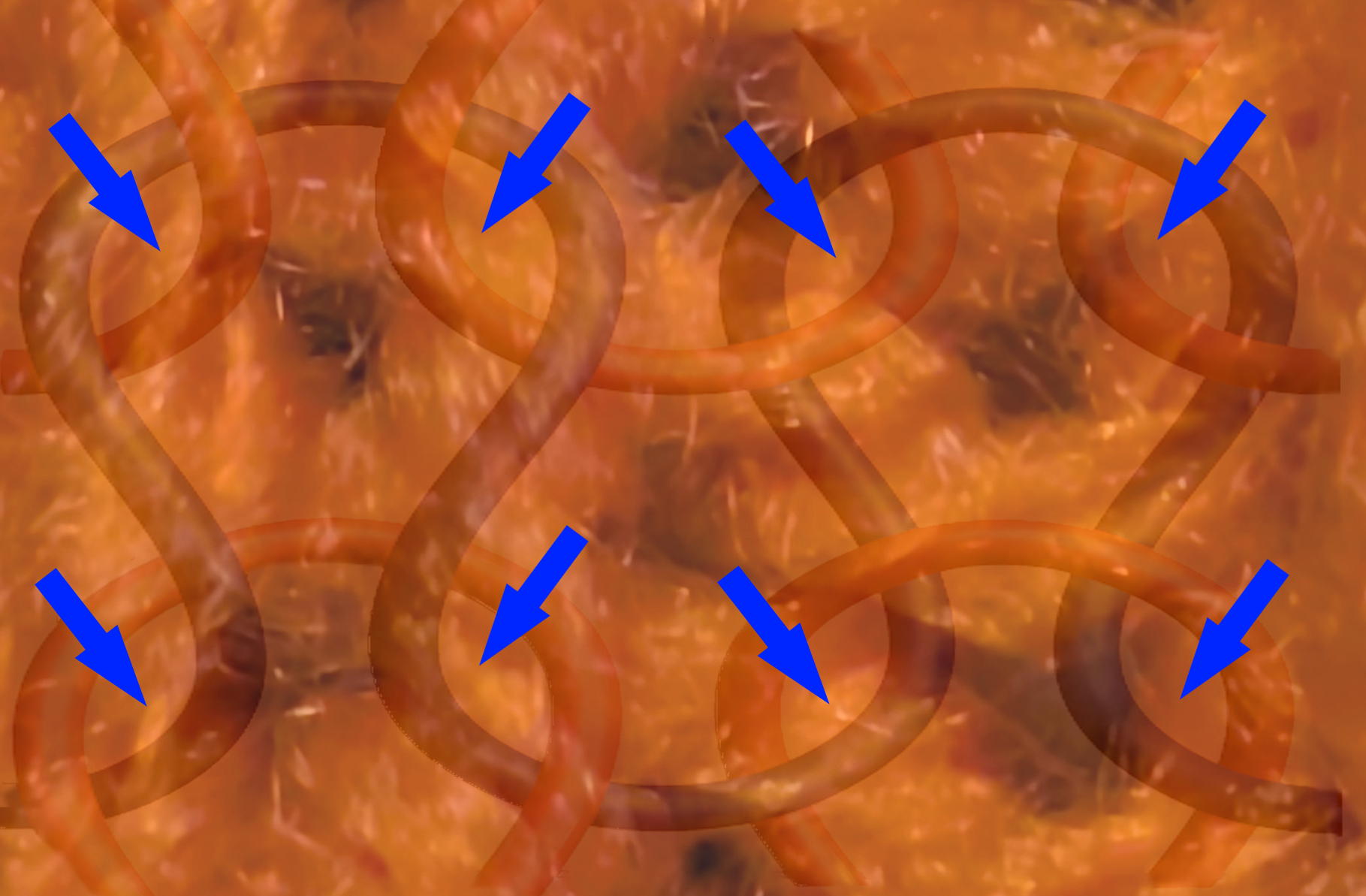}}	
    \put(5,5){\includegraphics[width=6cm,trim={0 1cm 0 1cm},clip]{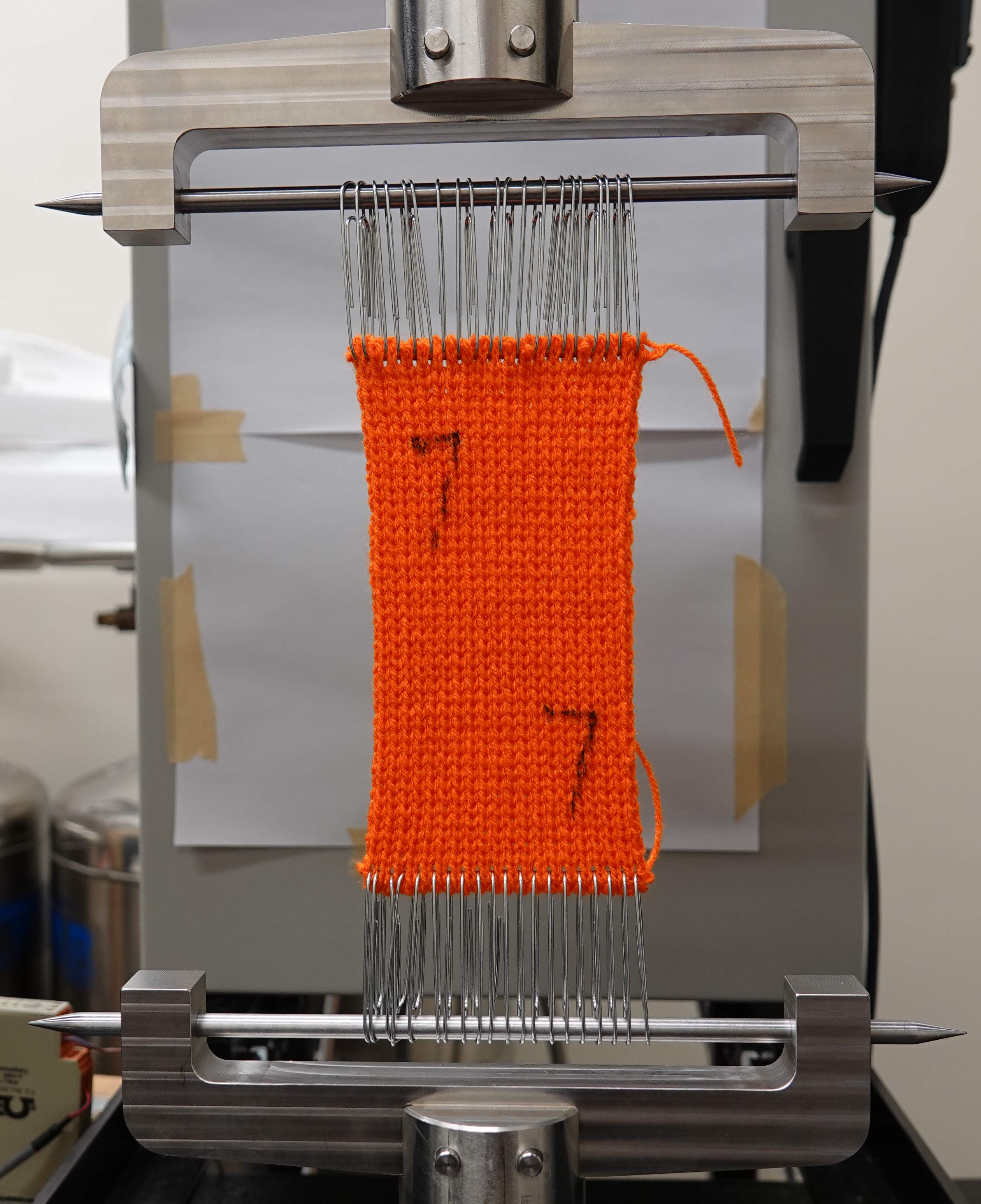}}
	\put(75,5){\includegraphics[width=6cm]{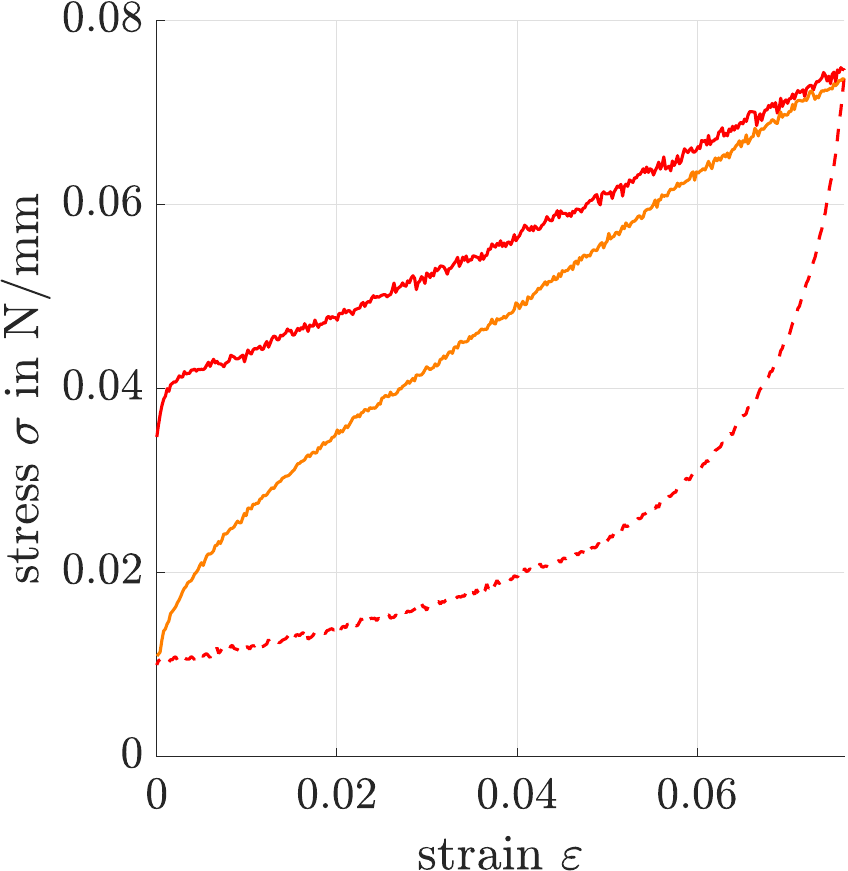}}
        \put(0,115){a)}
        \put(70,115){b)}
        \put(0,70){c)}
        \put(70,70){d)}
	\end{picture}
    \caption{
    {\bfseries a)} A ribbed fabric machine-knitted from acrylic fingering-weight yarn, stretched horizontally to show the full structure. The horizontal direction is the course and vertical is the wale. 
    {\bfseries b)} Zoomed-in fabric to show two unit cells, overlaid with a schematic of the yarns. Entangled regions are identified with blue arrows. Overlay provided by Michael Dimitriyev (private communication, October 2023) 
    {\bfseries c)} Experimental setup. The fabric is attached to custom made clamps with paperclips. The length of the fabric in the zero strain state, defined in App.~\ref{sec:dstar}, is $131\,\textrm{mm}$. The bottom clamp is connected to the base and the top clamp to the crosshead of an Instron universal testing machine, with which we take the sample through repeated load (crosshead moves up)--unload (crosshead moves down) cycles. More information about the sample fabrication and setup are detailed in App.~\ref{sec:fabrication} and App.~\ref{sec:instrumentation}. 
    {\bfseries d)} First loading (red solid line), first unloading (red dashed line), together with second loading curves (orange solid line) show significant hysteresis in the fabric response.
   }
    \label{fig:experiments}
\end{figure*}
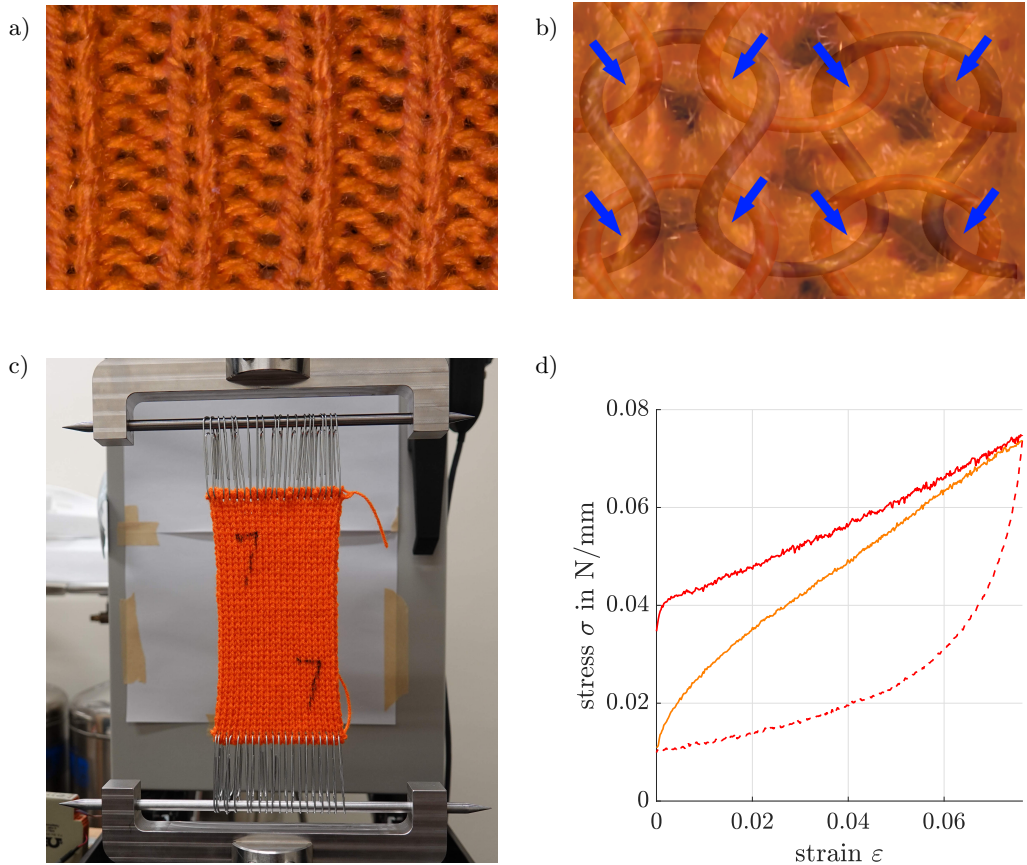

\begin{figure*}[t]
	\centering
	\unitlength=1mm
	\begin{picture}(140,90)
        \put(0,48){\includegraphics[width=6cm]{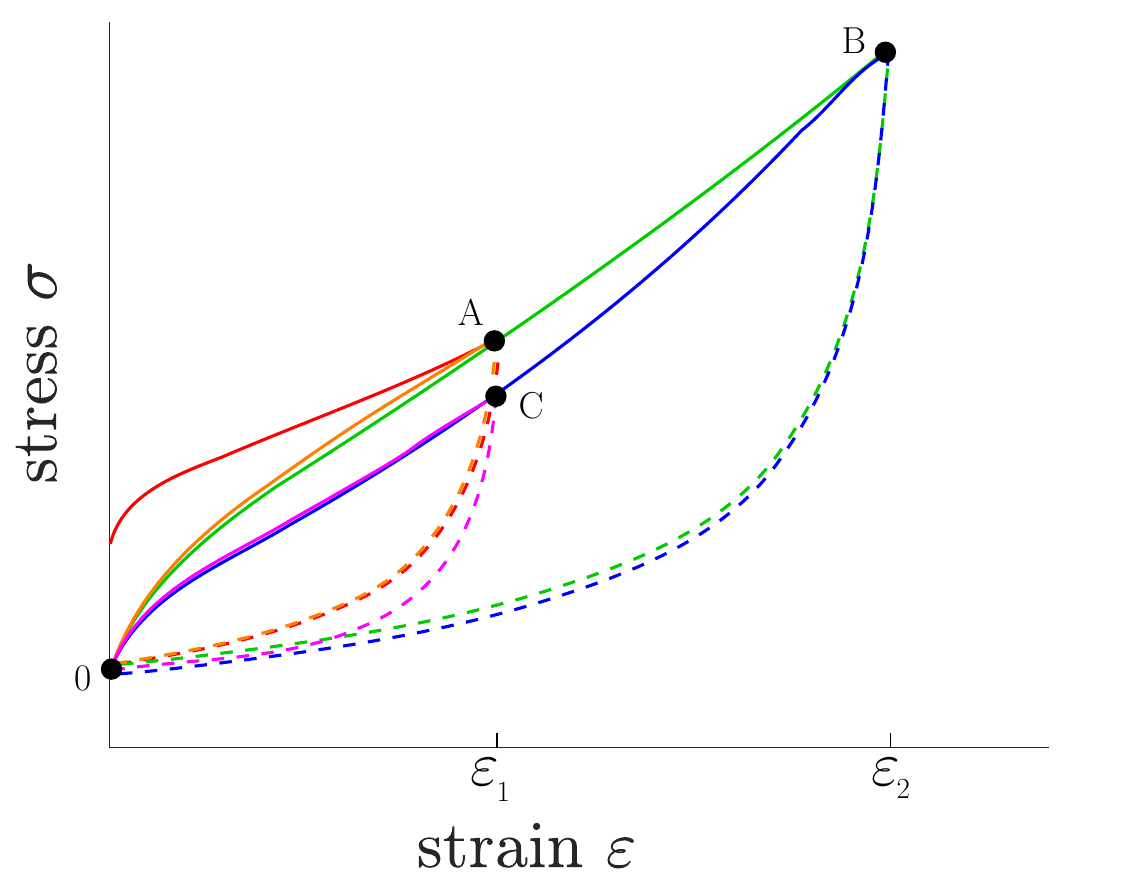}}
        \put(0,0){\includegraphics[width=6cm]{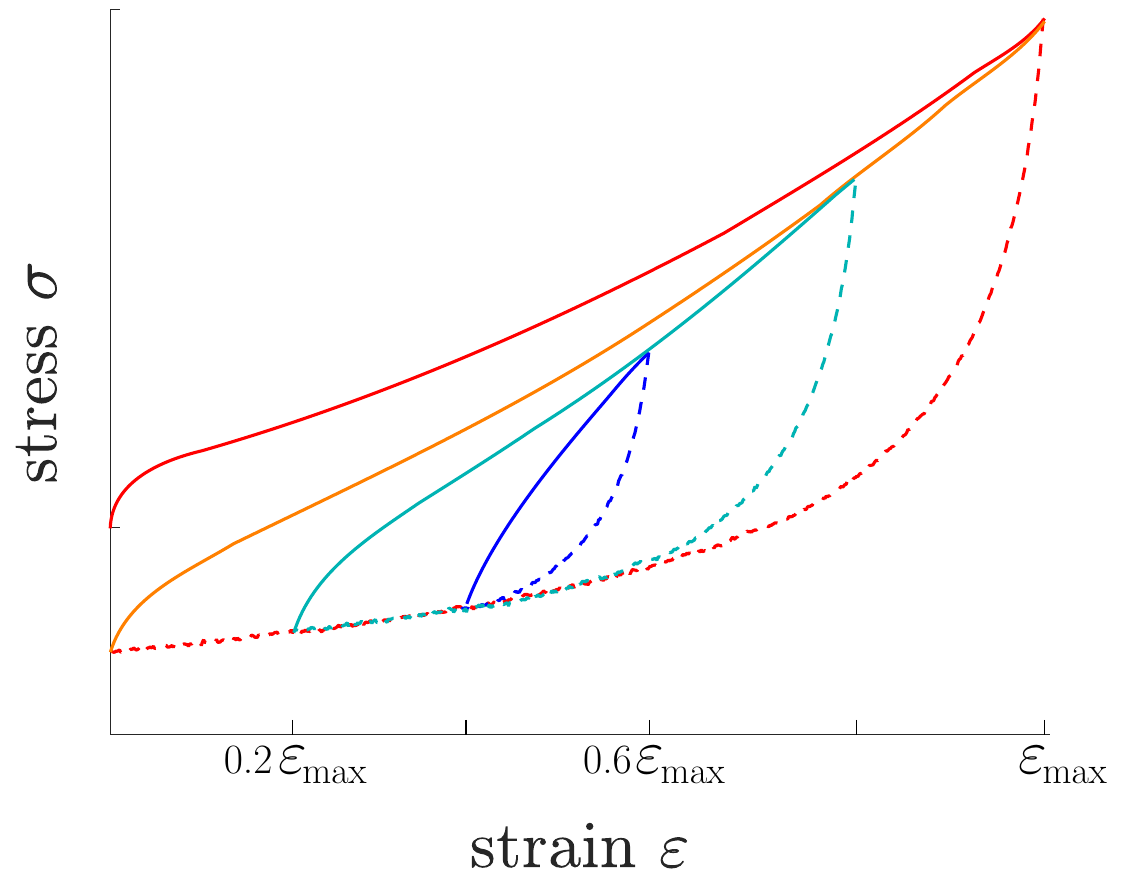}}
	    \put(70,48){\includegraphics[width=6cm]{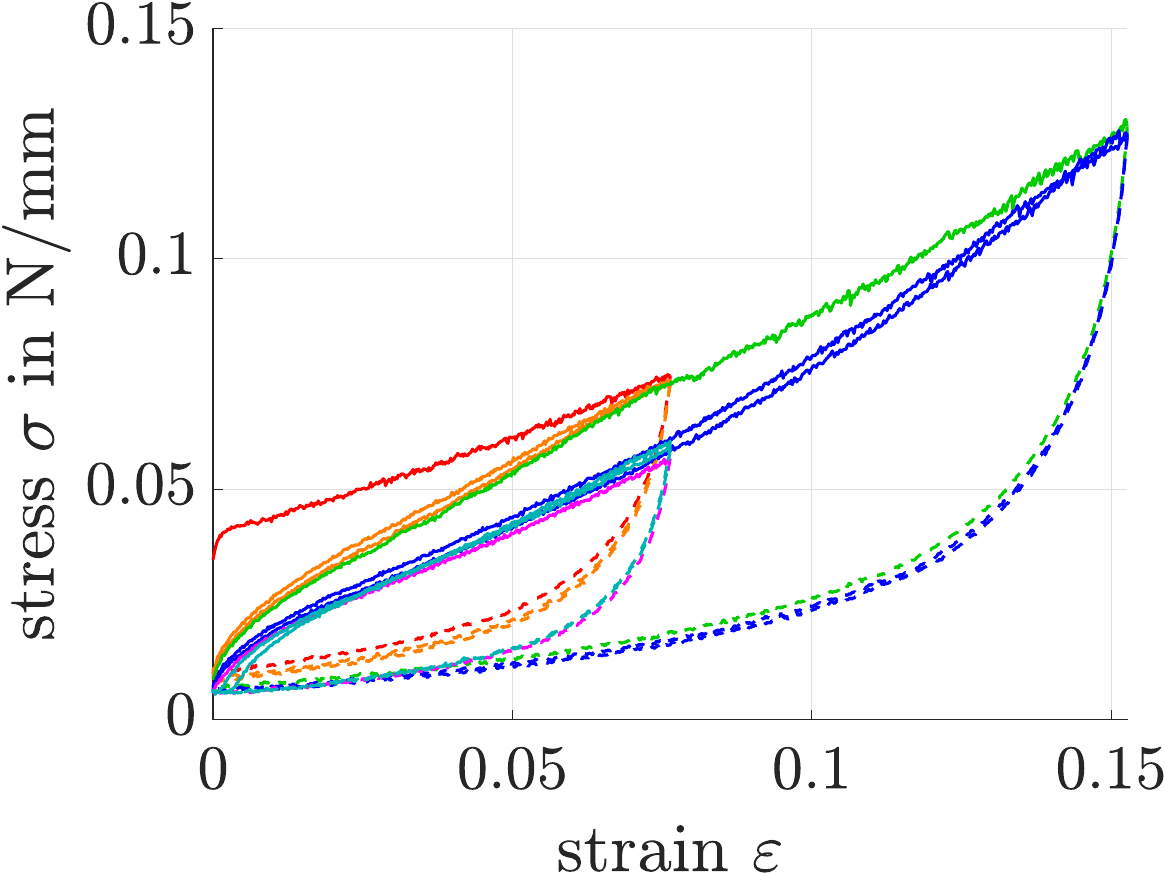}}
        \put(70,0){\includegraphics[width=6cm]{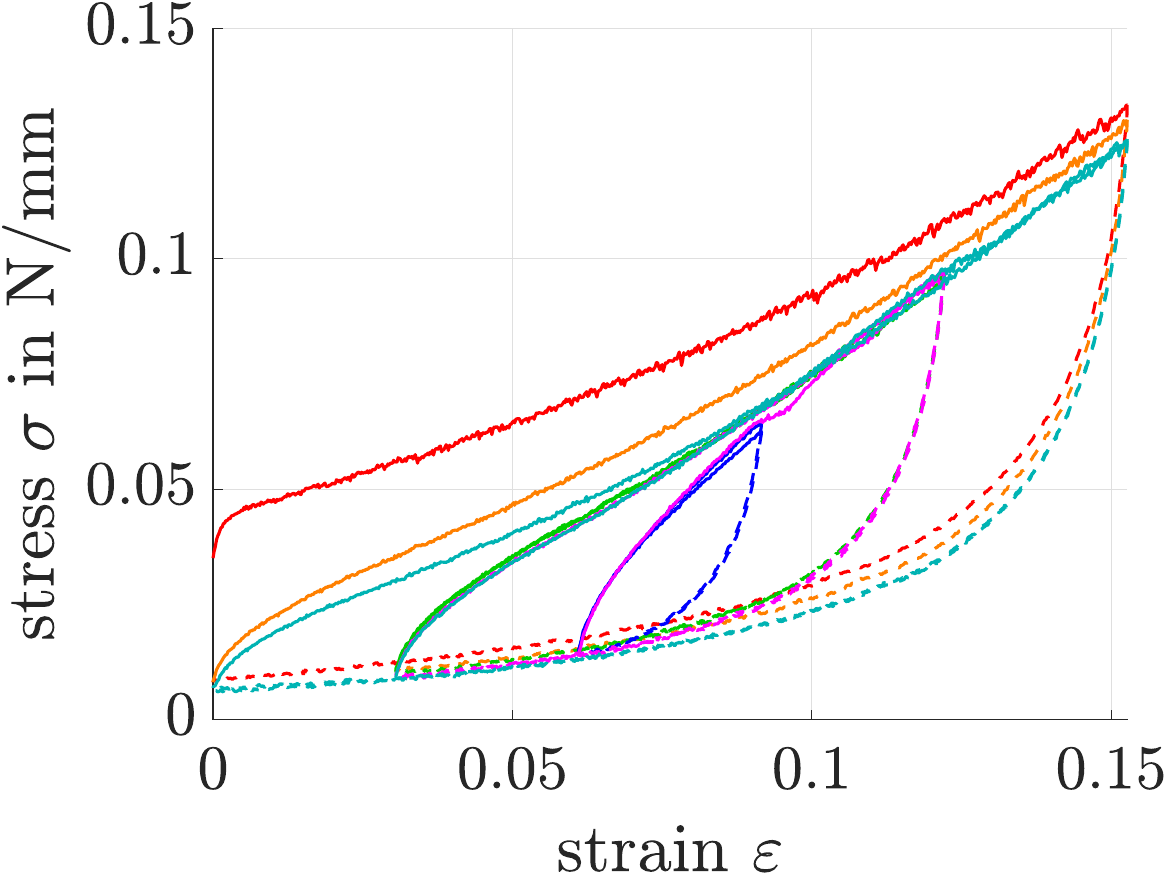}}
        \put(0,89){a)}
        \put(66,89){b)}
        \put(0,41){c)}
        \put(66,41){d)}
	\end{picture}
    \caption{
    {\bfseries a)} Idealized return point memory for a fabric first cycled to a given strain~$\varepsilon_1$ then subsequently to twice that strain, i.e.,~$\varepsilon_2$. The first cycle to the lower strain $\varepsilon_1 \sim 7.5\%$ is shown in red. The return point, $\sigma_1(\varepsilon_1)$, is denoted by $A$. Subsequent strain cycles to $\varepsilon_1$ follow the path $0 \longrightarrow A \longrightarrow 0$. The first cycle to higher strain $\varepsilon_2 \sim 15\%$ is shown in green and reaches a new return point $B$. Subsequent cycles to $\varepsilon_2$ follow the blue loading curve $0\longrightarrow C \longrightarrow B$ and the green unloading curve. When the fabric is subsequently cycled to $\varepsilon_1$ again, the loading reaches return point $C$. 
    {\bfseries b)} Experimental data for these strain cycles, first cycle in red, second in orange, and henceforth. 
    {\bfseries c)} When the fabric is first cycled from zero strain to $\varepsilon_\textrm{max}$ (red then orange curves), subsequent cycles to intermediate strains form nested loops, which rejoin the larger loop only at the extrema, within the larger loop (teal and blue curves). 
    {\bfseries d)} Measured nested hysteresis curves, including the reconnection paths between nested loops. The measurements were performed at $\dot{\Delta} = 0.25$ mm/s for a fabric of initial length $L = 131$ mm.}
    \label{fig:RPM}
\end{figure*}

While knitters have possessed an intuitive understanding of these materials for centuries, the formal scientific investigation of their emergent mechanics is a recent development \cite{Singal2023, Gonzalez2025JammedKnits, niu2025geometric, Ding2024KnittedMechanics, poincloux-crackling-2018, poincloux2018}. 
Key open questions include how constituent yarn properties dictate macroscopic behavior \cite{bini:01,santos:23,salopek:24} and the complete characterization of the fabric's emergent mechanical response.
 In studies that perform tensile tests on these materials, typically only a single loading phase is considered, although significant hysteresis has been reported in the corresponding unload phase in \cite{Matsuo2009HysteresisKnitted, poincloux2018}. The behavior of knitted fabrics under repeated cycles of deformation, crucial to a multitude of applications, has not yet been systematically studied.
In this work, we perform experiments on knitted fabrics to address their response under uniaxial loading and unloading; our experimental setup as presented in Fig.~\ref{fig:experiments} allows us to apply uniaxial stress cycles to knitted fabrics. Our results, shown in Fig.~\ref{fig:RPM}, reveal that these materials exhibit substantial hysteresis. Surprisingly, knitted fabric also exhibits return point memory: a fabric remembers the load at its highest extension when it undergoes repeated cycles of deformation. 

\section{Return point memory}
Return point memory has previously been observed in ferromagnets and predicted in anti-ferromagnets \cite{jiles1984,sethna1993,deutsch2004} and recently has been seen in sheared amorphous solids \cite{Fiocco2014EncodingMemory, Keim2020GlobalMemory} and unfolded crumpled sheets \cite{Shohat2022MemoryCrumpledSheets}. 
This phenomenon describes a state where the response (magnetization or stress)
 of a material under cyclic input (magnetic field or deformation) is dependent on the maximum value of the input in the material's history. The material response at this input, referred to as the return point, becomes cycle-number-independent. A unifying framework for the physical mechanisms of memory in materials is a frontier in modern physics \cite{Keim2019MemoryMatter} and knitted fabrics offer a wide and easily accessible landscape to explore memory in materials. 
To this end, in this work we elucidate a physical explanation of the observed memory properties of fabrics using the Preisach's model for ferromagnets \cite{preisach:35}, that captures the interplay between the underlying yarn and the knitted structure and well predicts our experimental results.

In Figs.\ \ref{fig:RPM}a,c, we show schematics of the idealized return point memory effect alongside the measured data (Figs.\ \ref{fig:RPM}b,d) for two classes of tensile measurements on pre-stressed fabrics. In the first class of measurements Fig.\ \ref{fig:RPM}b, we perform cycles between a reference zero strain (systematically defined in App.~\ref{sec:dstar}) and $\varepsilon_1$ corresponding to a $10\,\textrm{mm}$ displacement followed by cycles between zero strain and $\varepsilon_2 = 2\,\varepsilon_1$, followed by repeat cycles between zero strain and $\varepsilon_1$. 
The stress recorded after a cycle to a maximum strain is the return point (A, B labeled in Fig.~\ref{fig:RPM}a.   The loading to A sets the hysteresis cycles for all loading below $\varepsilon_1$.  Exceeding $\varepsilon_1$  creates a new return point B, which resets all hysteresis cycles for loading below $\varepsilon_2$.
The measurements reveal that memory in these materials is non-local in time and can be wiped out by taking the material to higher maximum strain. In the idealized schematics of these memory measurements, the return point is learned in a single cycle: this is nearly matched in the measurements.

In the second class of measurements, schematically shown in Fig.\ \ref{fig:RPM}c, we consider cycles between zero strain and $\varepsilon_\textrm{max}$ followed by cycles between intermediate strains. This sequence of loading/unloading reveals the congruency effect: cycles between intermediate strains form nested hysteresis loops whose size and shape is determined solely by the return point established at $\varepsilon_\textrm{max}$, the largest strain in the material's history.  Fabrics exhibit the congruency behavior as seen in Fig.\ \ref{fig:RPM}d.
This congruency corresponds to the current return point, making these fabrics distinct from the congruency effects of nested hysteresis loops observed in phase transforming materials such as shape memory alloys \cite{RAO20144554}.

\begin{figure*}[t]
	\centering
	\unitlength=1mm
	\begin{picture}(160,70)
    \put(-2,0){\includegraphics[width=5.2cm]{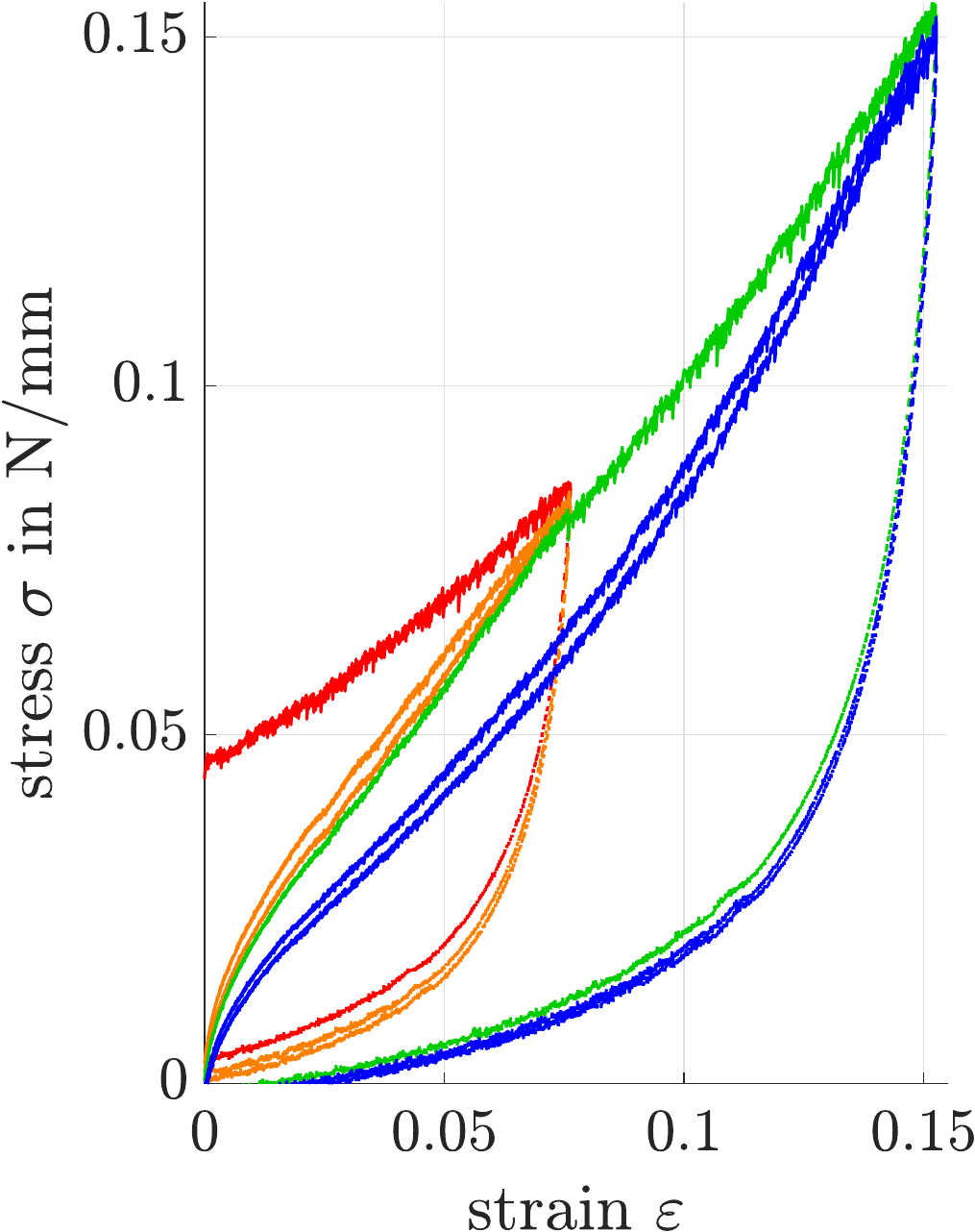}}
    \put(51,0){\includegraphics[width=5.2cm]{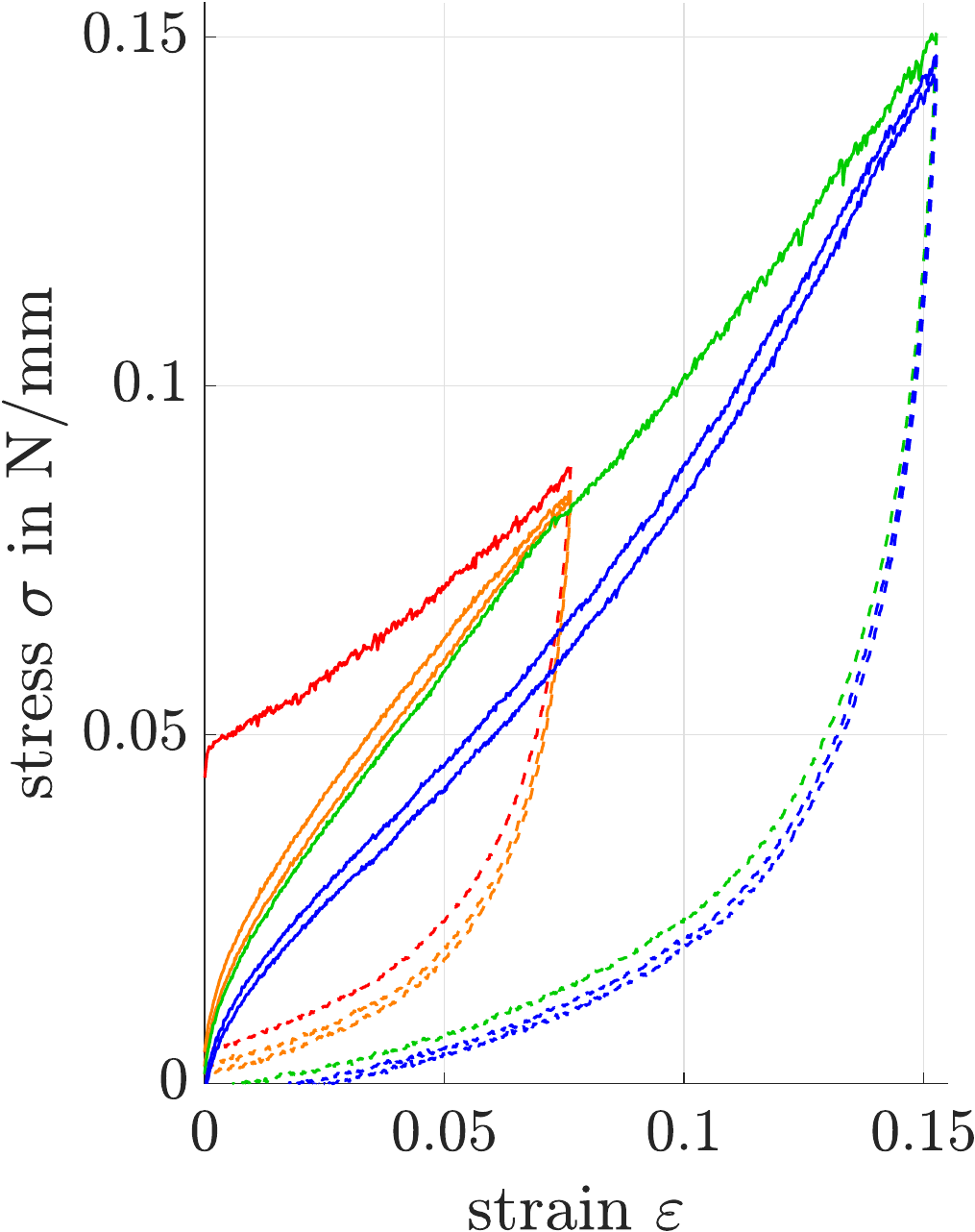}}
    \put(104,0){\includegraphics[width=5.2cm]{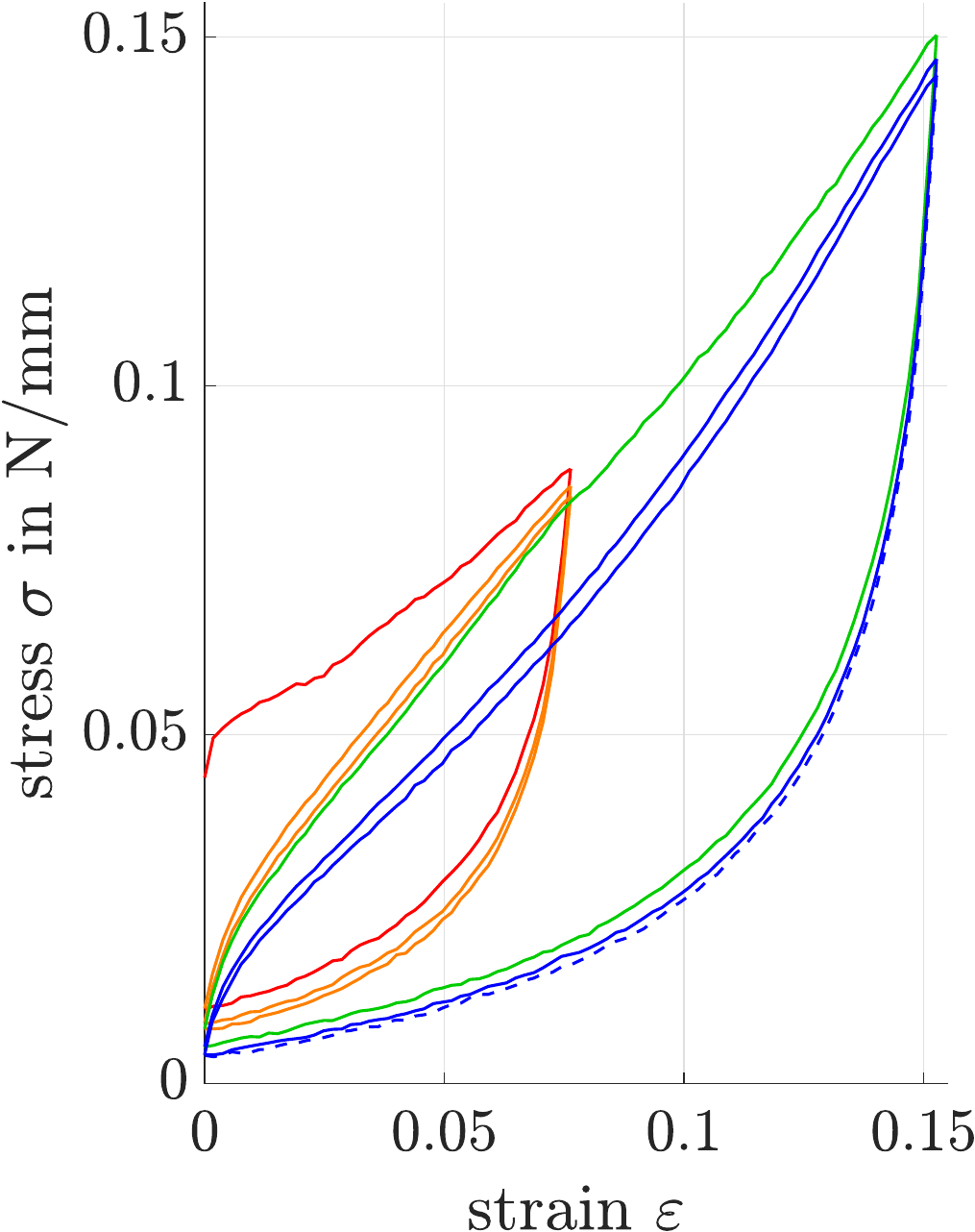}}
    \put(0,68){a)}
    \put(53,68){b)}
    \put(106,68){c)}
	\end{picture}
    \caption{ 
    Measurements of cycles between zero strain (systematically defined in App.~\ref{sec:dstar}) and $\varepsilon_1\approx 0.075$ followed by cycles to $\varepsilon_2\approx 0.15$ done at three different strain rates corresponding to 
    {\bfseries a)}~$\dot\Delta = 0.025\textrm{mm/s}$, 
    {\bfseries b)}~$\dot\Delta = 0.25\textrm{mm/s}$ and 
    {\bfseries c)}~$\dot\Delta = 2.5\textrm{mm/s}$. 
    The fabric's initial length was $L = 135$ mm. Between each measurement, the fabric was taken off the universal testing machine and rested horizontally on a table for ten minutes. These measurements were done on the same sample, which was identically prepared to the sample measured in Figs.\  \ref{fig:experiments} and \ref{fig:RPM}.}
    \label{fig:rate}
\end{figure*}
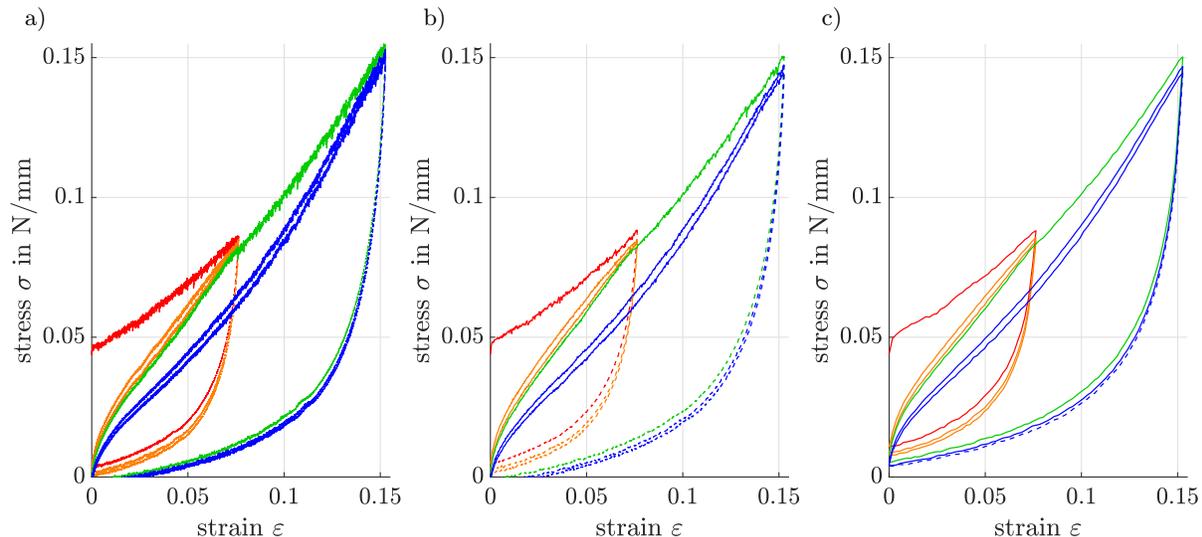

The memory properties of the fabric are reasonably independent of strain rate. 
Measurements performed at extension rates spanning two orders of magnitude $(\dot\Delta = 0.025~\textrm{mm/s}$, $0.25~\textrm{mm/s}$, and $2.5~\textrm{mm/s})$ -- roughly corresponding to strain rates of $2\times 10^{-4}~\textrm{sec}^{-1}$, $2\times 10^{-3}~\textrm{sec}^{-1}$, and $2\times 10^{-2}~\textrm{sec}^{-1}$ -- reveal that the memory properties are effectively rate-independent (Fig.\ \ref{fig:rate}); see App.~\ref{sec:relax} for additional details. Accordingly, we  perform all experiments in this work at extension rate $\dot\Delta = 0.25~\textrm{mm/s}$ and assume rate-independence in our model. The stretch direction and topology of the underlying fabric are also not relevant to obtaining the qualitative features of return point memory, as shown in App.~\ref{sec: other_fabrics}. 

% \section{Figure 3}
% 1.  Rate independence (data)

%\section{Figure 4}
%1.  Cartoon?  Probably not.
%
%2.  Model response on top of data of Figure 2

\section{Model}
The rate-independent hysteresis and return point memory behaviors of fabrics rule out conventional modeling via viscoelasticity or plasticity, necessitating a distinct microscopic picture of the phenomenology seen.
The Preisach model of hysteresis \cite{preisach:35, mayergoyz:86,krasnoselskii.pokrovskii:89} which captures the features of non-local memory, the wiping-out property, and the congruency property of return point memory \cite{sethna1993} provides a convenient framework. 
The elementary unit of the Preisach model, known as a hysteron, inhabits two possible states, typically referred to as `on' and `off'. By considering an ensemble of non-interacting hysterons with differing threshold values for switching from `on' to `off' and vice versa, one can reproduce general hysteretic behaviors. In knitted fabrics, contacts between yarns in entangled regions of the knitted structure, illustrated in Fig.\ \ref{fig:experiments}a, play the role of hysterons. As the fabric is stretched, yarns in these regions compress radially. During the unloading phase, the yarn in these contact regions has been compressed and thus does not resist deformation with the same strength that it did during the loading phase. 

As shown in Fig.~\ref{fig:RPM}, there are strong nonlinearities in the hysteresis which must be incorporated into the  Preisach model.  In particular, fabrics show hysteresis that is asymmetric and  evolves with the maximal strain seen by the fabric.
Mathematically, this requires the  Preisach hysteron distribution function, $\mu$, to depend on a measure of entanglement, 
$\varepsilon_t$, a strain measure that evolves asymmetrically during loading cycles,
and the maximal strain up to the current time~$t$, $\varepsilon_\mathrm{max} = \max_{t'\le t}\varepsilon[t']$.
Likewise the Preisach relay function, $s$, must also depend on these same measures of material state.
Precisely, we write the stress at time $t$ as
\begin{widetext}
\begin{equation}
	\sigma(\varepsilon[t],\varepsilon_t,\varepsilon_\mathrm{max}; \alpha, \beta) =\int\int_{\alpha\geq\beta}\mu(\alpha',\beta',\varepsilon_t,\varepsilon_\mathrm{max})\,s(\varepsilon[t],\varepsilon_t,\varepsilon_\mathrm{max}; \alpha', \beta')\,\mathrm{d}\alpha'\,\mathrm{d}\beta'\,,
\end{equation}
\end{widetext}
with Preisach density function~$\mu(\alpha,\beta,\varepsilon_t,\varepsilon_\mathrm{max})$, lower and upper thresholds~$\beta\leq\alpha$, and relay function $s(\varepsilon[t],\varepsilon_t,\varepsilon_\mathrm{max}; \alpha, \beta)$ for total strain~$\varepsilon$.

While the Preisach density function $\mu$ can theoretically be identified from experimental data via numerous loading-unloading cycles, the complex, nonlinear behavior of knitted fabrics renders analytical identification intractable; see supplement Fig.~\ref{fig:preisach_a}.  Notwithstanding, even if the hysteron density and the relay function can be identified, one is faced with the need to perform inconvenient integrations over the hysteron phase-space.
These points motivate the introduction of an alternate means to estimate the response of the Preisach model that simplifies the identification process and simultaneously facilitates the numerical evaluation of the model.

\begin{figure*}[t]
	\centering
	\unitlength=1mm
	\begin{picture}(150,58)
		\put(-5,0){\includegraphics[width=0.45\textwidth]{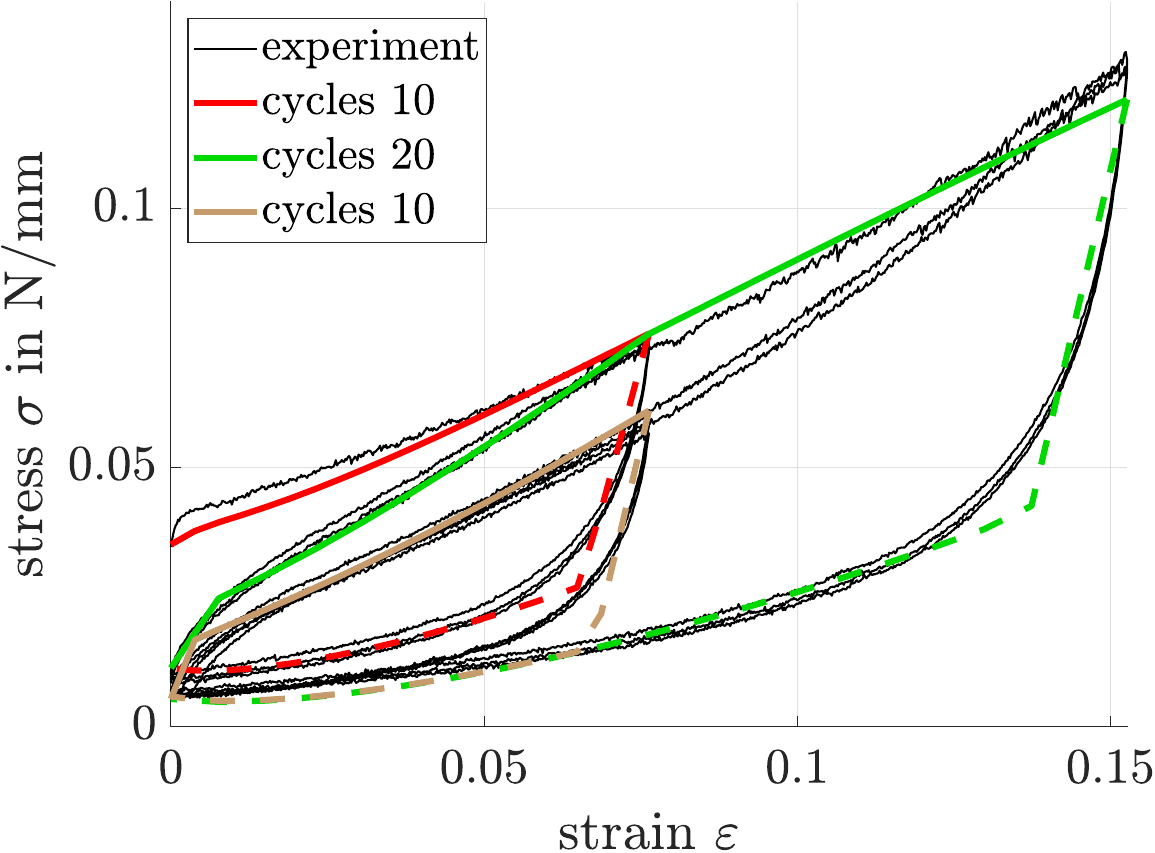}}		
		\put(80,0){\includegraphics[width=0.45\textwidth]{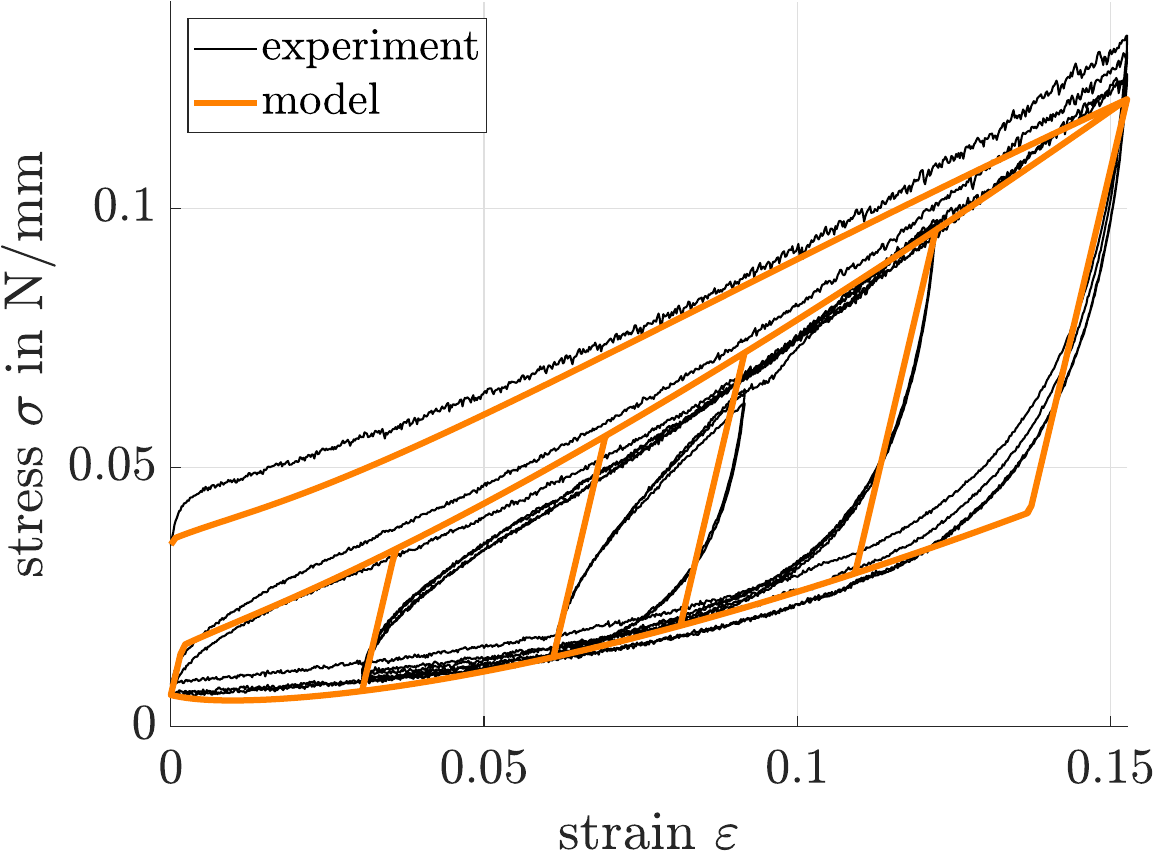}}
        \put(0,55){a)}
        \put(80,55){b)}
	\end{picture}
    \caption{
    {\bfseries a)} Numerical simulation of a tensile test of a knitted fabric utilizing an extended Preisach model reproduces the experimental results very well. 
    {\bfseries b)} Taking into account the nested looping confirms that the three key features - non-local memory, wiping-out and congruency - are captured in the model. Note the initial stress of the simulation is set a priori to the initial stress from the experiments.}
    \label{fig:preisach_model}
\end{figure*}

We decompose the total strain into elastic and entanglement components, $\varepsilon=\varepsilon_e + \varepsilon_t$.  Stress is defined as proportional to the elastic strain~$\sigma=E\,\varepsilon_e$ and modulus~$E(\varepsilon_\mathrm{max})$ -- a function of the maximal strain history~$\varepsilon_\mathrm{max}$. 
The entanglement component only evolves when the stress hits the rebound boundary, which can be directly inferred from the hysteresis loops similar to the established Preisach model. Specifically, the history dependent rebound inequality $g(\sigma,\varepsilon_t,\varepsilon_\mathrm{max})=|\sigma - R_C(\varepsilon_t,\varepsilon_\mathrm{max})| - R(\varepsilon_t,\varepsilon_\mathrm{max}) < 0$ defines material states where $\dot\varepsilon_t = 0$.  $R$ defines the rebound range and $R_C$ locates the center of the range -- analogous to a kinematic plasticity model of dislocation accumulation at a grain boundary \cite{anandgovindjee}.  The evolution of the entanglement strain, $\dot\varepsilon_t = \gamma\, \partial g/\partial \sigma$, is fully governed by the Karush-Kuhn-Tucker conditions $\gamma\ge0$, $g \le 0$, $\gamma g = 0$ \cite{simo98}.  Details of the functional forms are provided in the supplemental materials, Sec.~\ref{sec:supmodel}.

In Fig.\ \ref{fig:preisach_model}a,
we fit the model from the experimental data
of cycling three times to~$\varepsilon_1$,~$\varepsilon_2$ and~$\varepsilon_1$ again, cf. Fig.~\ref{fig:RPM}b. 
Figure~\ref{fig:preisach_model}a demonstrates the good fitting capabilities of the model, where the experimental data is shown in black and the model fit in red, green, and brown for cycles to $\varepsilon_1$,~$\varepsilon_2$, and again to $\varepsilon_1$, respectively.  
This very simple model is observed to be a suitable abstraction of the Preisach model to describe return point memory as well as the dependency of the return points on the maximal experienced strain. 
Using the fit model, we apply it to the nested hysteretic experiments as shown in Fig.~\ref{fig:RPM}d. The predictions are shown in Fig.~\ref{fig:preisach_model}b, where 
we see that the model also incorporates the return point memory wiping-out and congruency effects while still providing reasonable quantitative predictions.

\section{Discussion} 
Our experiments establish that knitted fabrics exhibit the classic features of return point memory.
By identifying yarn contacts within the knitted structure as hysterons, we developed a model that captures
 the wiping out, congruency, and temporal nonlocality properties of return point memory materials. The model well predicts the experiments and replicates, in a very convenient way, a Preisach model using an abstraction from kinematic-hardening in plasticity theory. 

Our results open new questions and possibilities for a wide range of textile materials. The yarn-contact interpretation of hysterons in these fabrics extends to general knitted fabrics as well as other loop-based fabrics constructed for example by crochet or hybrid crochet. Exploring the memory properties of these varieties of loop-based fabrics constructed from yarn is a fruitful question for future work. Return point memory may also be present in knot-based materials of interest beyond loop-based fabrics, such as woven fabrics and torus knot tessellations \cite{Yan2025RigidityChainmail}. The memory or lack thereof in these materials will be a useful exploration to pinpoint the relevant parameters for material memory. This understanding of material memory, in addition to theoretical interest, will give an additional facet of these materials that can be engineered in fabric applications. Interestingly,
the spatial symmetries of fabrics constrain the tensorial properties of these materials. In fabrics that fall into some wallpaper groups, anomalous tensors are allowed \cite{Dresselhaus2025AnomalousTensorial2D}. The interrelation between anomalous tensors and memory is an additional open question for future study.

Fabrics further offer a platform that is both simple to visualize and to tune for seeking other types of memory. For example, the development of multiple memories in a material through interacting hysterons \cite{Lindeman2021MultipleMemory} may be realized by knitting the fabric with a higher tension (see App.\ \ref{sec:fabrication}) such that yarn contacts are closer together. A comprehensive study of different constituent materials (from acrylic, cotton, and wool yarns to nylon fiber) knitted at different tensions could reveal the effect of hysteron interactions on fabric memory. In knitted fabrics it is also possible to engineer the spatial symmetry of hysteron interactions, a phenomenon not realizable in jammed packings. Spatially symmetric interactions can have observable consequences; for example, dielectric screening of optical phonons in graphene is dependent on the symmetries of the underlying hexagonal lattice \cite{Manes2007SymmetryElectronPhononGraphene, Moczko2025DielectricScreeningGraphene}. Experimental and theoretical exploration of the hysteron-interactions in stitch patterns with different spatial symmetries could open further avenues to tune memory properties through fabric design.

\section{Acknowledgments}
EJD, SG, and KKM acknowledge support from the National Science Foundation (Grant No.\ CMMI-2344589) and the College of Chemistry, UC Berkeley. Any opinions, findings, and conclusions expressed in this publication are those of the authors and do not necessarily reflect the views of the NSF. KKM also acknowledges partial support from the Director, Office of Science, Office of Basic Energy Sciences, of the U.S. Department of Energy under contract No. DEAC02-05CH11231.
Further, EJD thanks  A.P.\ Cachine, K.\ Singal, M.\ Dimitriyev and   E.\ Matsumoto for helpful discussions.
This research was partly funded by the UDE Postdoc Seed Funding (SH).

\bibliography{apssamp}
\clearpage

\appendix
\section{Supplementary Experimental Details}

\subsection{Sample preparation}\label{sec:fabrication}

We fabricate samples with the SilverReed SK840 knitting machine in conjunction with the SilverReed SRP 60N ribber. The key components and their functions are described in \cite{SilverReedSRPRibberInstruction}. All samples are knitted from 3-ply orange acrylic yarn of average diameter $0.95\,\textrm{mm}$, TAMM brand. We follow the protocol below to make ribbed samples. Steps 1-6 detail the creation of the cast-on and lower boundary in the wale direction. In step 7, the bulk of the fabric is created and in steps 8-12 the upper boundary in the wale direction and cast-off are created.
   
\begin{enumerate}\itemsep-0.8ex
\item Align 20 needles on the knitting machine bed and 20 needles on the ribber bed in an alternating arrangement. 
\item Position the carriage ensemble (knitting machine plus ribber carriage) to the left side of the knitting machine. Pass the carriage ensemble twice over the needle bed to properly align all 40 needles. 
\item Thread the carriage with the working yarn. 
\item Set the dials on both carriages to tension 3. Adjust the set lever to [0 1] and the cam lever to “stockinette”. 
\item Move the carriage to the opposite side to knit one row. Then attach the cast-on comb and add weights to hold the fabric in place. 
\item Switch the cam lever to “circular”. Keeping all other settings the same, knit 4 additional rows. 
\item Change the cam lever back to “stockinette”, adjust the set lever to [1 1], and set the tension to 5. Tension 5 is slightly looser than tension 3, ensuring roughly even tension between the cast-on and bulk of the fabric. Knit 40 rows. 
\item Set the cam lever to “circular”, set lever to [0 1], and reduce the tension to 3. Knit 4 rows. 
\item Cut the working yarn, leaving a sufficient length for casting off later. Thread the carriage with a contrasting yarn. 
\item Set the tension to 3, cam lever to “stockinette”, and set lever to [1 1]. Knit 6 to 8 rows with the contrasting yarn. 
\item Cut the yarn and move the carriage to the opposite side. The knitted fabric will detach from the machine. Remove the cast-on comb and weights. 
\item Cast off the fabric using the loose end of the original yarn, sewing all adjacent stitches of the final row together. Steps 8-12 ensure symmetry between the cast-on and cast-off. Remove the contrasting yarn after casting off is complete. \end{enumerate}
The ribbed fabric sample is now complete and ready for use. Typical rest dimensions for these fabrics are approximately $85\,\textrm{mm}$ (along course) by $135\,\textrm{mm}$
(along wale).

\begin{figure}[t]
    \centering
    \includegraphics[width=0.70\linewidth]{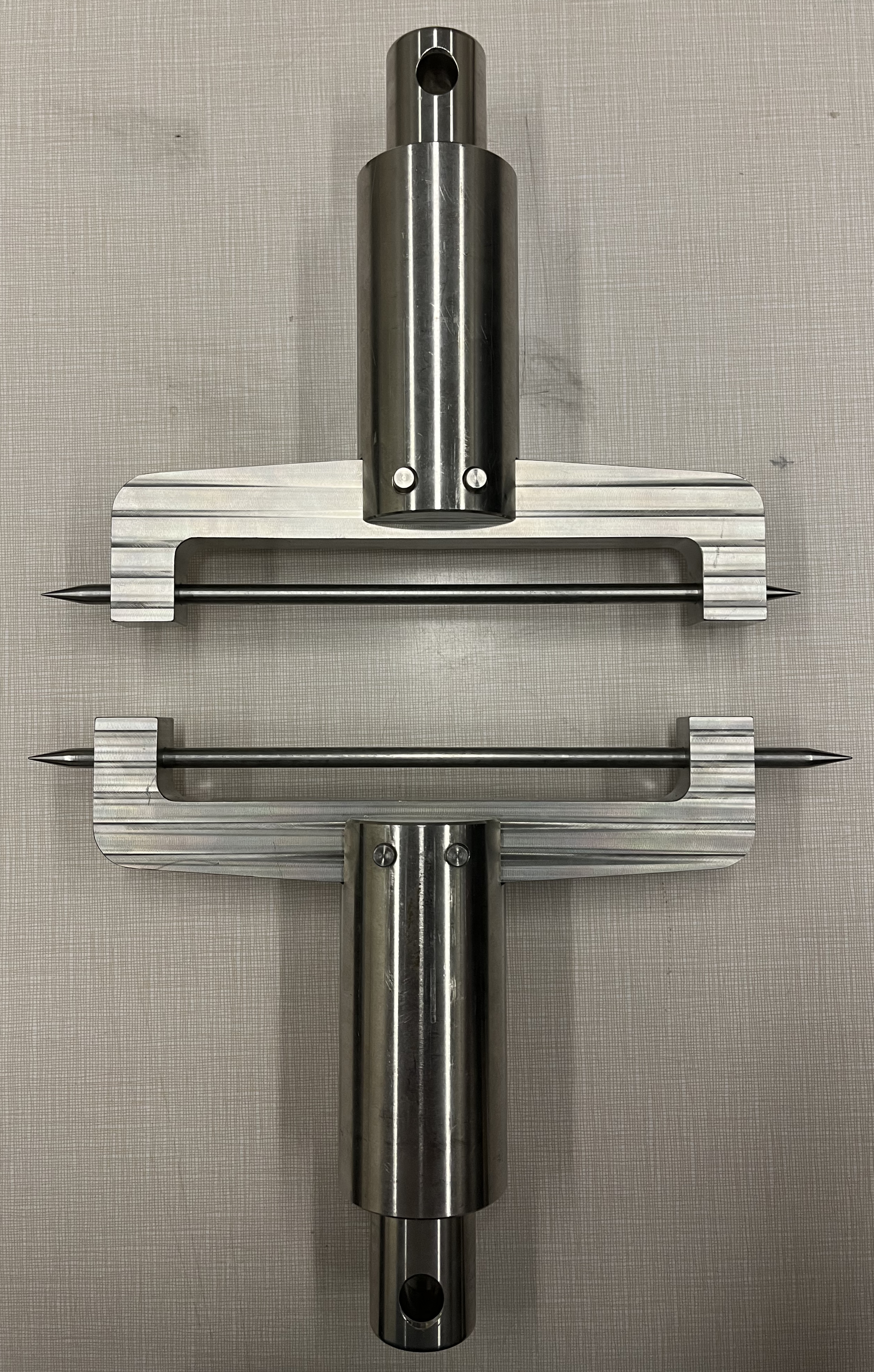}
    \caption{Custom clamps that were attached to the tensile testing machine.}
    \label{fig:clamps}
\end{figure}

\subsection{Instrumentation Details}\label{sec:instrumentation}
All measurements were conducted using an Instron Dual Column Model 5965 testing machine. Below we provide specifications about the machine's frame and load cell:
\begin{itemize}\itemsep-0.8ex
    \item frame model: 5965 (tabletop)
    \item max. load capacity: $5\,\textrm{kN}$
    \item max. head of crosshead: $3000\,\textrm{mm/min}$
    \item strain accuracy: $\pm0.5\%$ of reading to $1/100$ of full scale with extensometer
    \item position accuracy: $\pm0.01\,\textrm{mm}$ or $0.05\%$ of displacement (whichever is greater)
    \item load cell model: 2580 series
    \item capacity: $5\,\textrm{kN}$
    \item accuracy: \\
    measurements of $10\,\textrm{N}$: $\pm0.5\%$ of reading or $\pm5\text{g}$\\
    loads between $20\,\textrm{N}$ and $50\,\textrm{N}$: $\pm0.5\%$ of reading\\
    loads between $50\,\textrm{N}$ and $5\,\textrm{kN}$: $\pm0.4\%$ of reading\\

\end{itemize}

\begin{figure*}[t]
	\centering	
    \includegraphics[width=\textwidth]{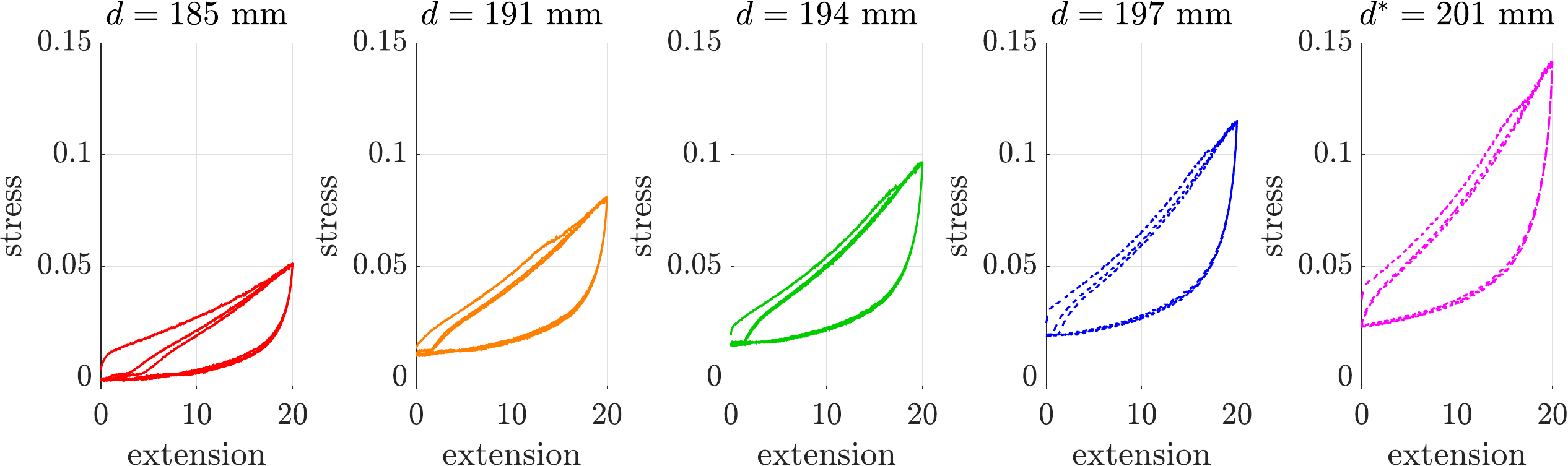}	
    \caption{Stress-extension data from each step of the protocol to find $d^*$. The uppermost curve on each plot is the first loading cycle. In the first (left most) test, the clamps are set a distance $d = 18.5\,\textrm{cm}$ apart, corresponding to a fabric length $L = 11.5\,\textrm{cm}$. On the first and second cycles of strain, $3$~mm of slack develop, seen by no rise in load after zero extension in the second and third cycles. To offset this slack, we raise the top clamp by $6\,\textrm{mm}$ for the second test (orange plot) and repeat the test. We tune $d$ by raising the clamp until no slack is accrued, as in the rightmost figure. We then set this clamp distance to $d^*$ and begin measurements from this configuration after taking the sample off of the Instron for ten minutes or longer. The measurements were performed at $\dot{\Delta} = 0.25$ mm/s.}
    \label{fig:d*}
\end{figure*}

We control the machine with BlueHill software, version 3.13. This software moves the crosshead at a user-given rate between user-given extensions.

To connect the samples to the Instron machine for uniaxial tensile testing measurements, we use custom-made clamps along with paperclips. The clamps (Fig.~\ref{fig:clamps}) were designed to accommodate transverse contraction caused by the Poisson effect during tensile testing, to ensure that the stress on the sample remains uniaxial.

To attach the sample to the clamps, we use paperclips (silver jumbo, Office Depot brand, manufacturer 10004BX item 429175). One end of each paperclip is threaded through the needle piece of the clamp (see Fig.~\ref{fig:clamps}) and the other is threaded through the fabric. To uniformly apply the load across the sample, these paperclips are threaded one per unit cell of the fabric, and are always threaded into the third row of the fabric (leaving three strands of yarn on each paperclip). Since each fabric sample has four rows of boundary fabric surrounding each side of the bulk fabric, the paperclips do not directly touch the bulk fabric. All lengths of the fabric in this work are measured between the closer tips of aligned paperclips on the top and bottom of the sample. 

Paperclips were chosen for this purpose because of their ability to move laterally along the needle during measurements of the fabric. This freedom of movement helps prevent stress concentrations and minimizes potential distortions in the fabric’s response to applied loads. This is also a controllable, inexpensive and fast method to perform repeated measurements on a variety of samples.

\begin{figure*}[t]
	\centering	
	\unitlength=1mm
	\begin{picture}(130,60)
		\put(0,0){\includegraphics[width=.8\textwidth]{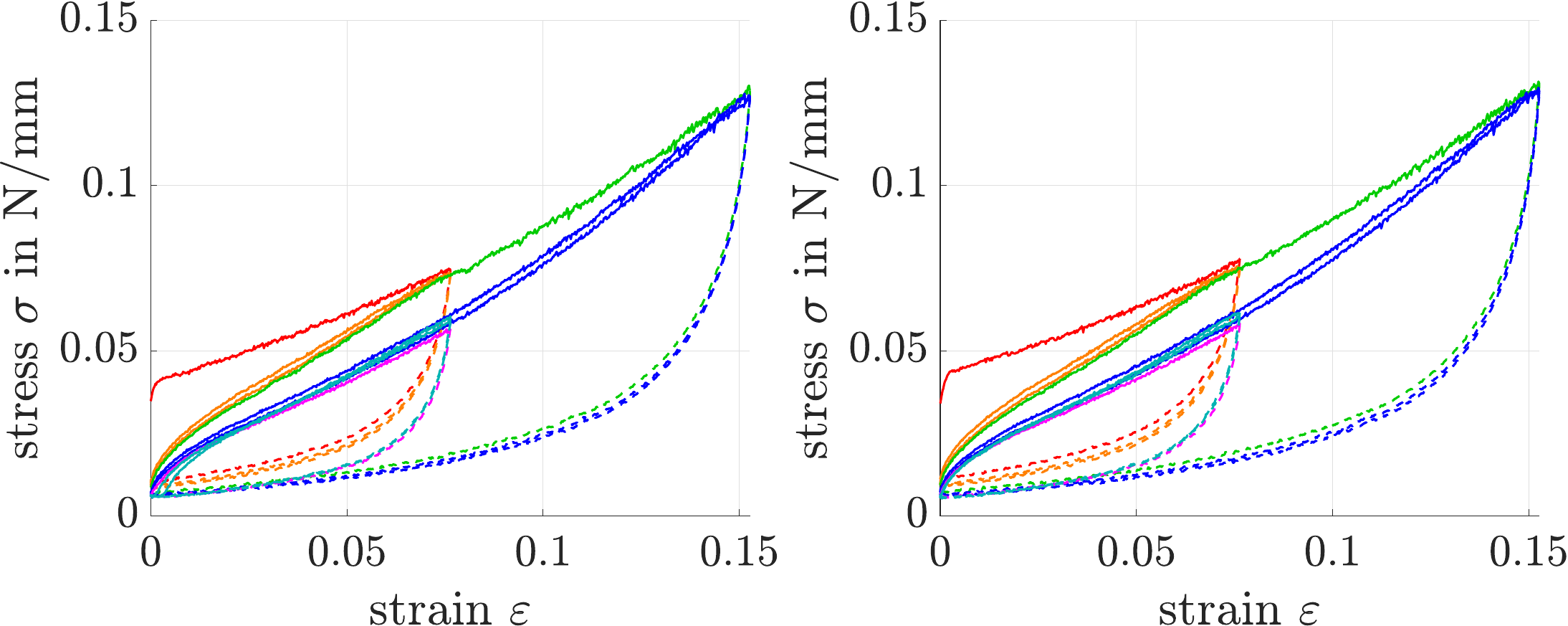}}
        \put(-2,55){a)}
        \put(70,55){b)}
	\end{picture}    		
    \caption{Return point memory demonstration for ribbed fabric stretched in the wale direction. {\bfseries a)}  The fabric is first stretched to $\varepsilon_1 \sim .075$ (red, then, orange curves), then to $\varepsilon_2 \sim .15$ (green, blue) and then again to $\varepsilon_1$ (pink, cyan). The measurements in {\bfseries b)} are taken on the same sample one day after those in a). The measurements were performed at $\dot{\Delta} = .25$ mm/s for a fabric of initial length $L = 131$ mm.}
    \label{fig:nextday}
\end{figure*}

\subsection{Demonstration of repeatable experiments}\label{sec:dstar}

In order to measure data in a reproducible manner we systematically determine the length of the fabric in the Instron Machine that corresponds to what we denote as the zero strain state. This state is standardized by setting a fixed distance between the upper and lower clamps at the start of each measurement. To find this fixed distance, which we denote as $d^*$, we perform a set of tensile tests that cycle 3 times to the maximum displacement planned for the measurement, increasing the distance between clamps until the second and third cycles show an immediate change in load at the start of the loading phase. We begin this protocol with clamp distance that gives a relatively slack fabric configuration, minimizing any residual strain in the system. This reference configuration for the conditioning protocol is defined by setting the distance between the clamps equal to the natural length of the fabric–paperclip ensemble. From this baseline, the fabric is subjected to repeated cyclic loading and unloading. As illustrated in Fig.~\ref{fig:d*}, at the beginning of the protocol, these cyclic tests reveal a progressive development of slack in the force-displacement curves at the beginning of the second and third strain cycles. For the next test, we raise the clamp by approximately $1.5-2$ times the slack acquired during the test. Over repeated tests and modifications, this slack converges to zero and a limiting configuration for the fabric. This limiting distance is denoted as $d^*$, and it marks the point at which the fabric exhibits a repeatable mechanical behavior under further cycling. We refer to this configuration as the zero-strain state and use it for all measurements of a fabric to the maximum displacement used in the protocol. We note that $d^*$ must be determined separately for all fabrics and maximum displacements to which they will be subjected in subsequent measurements.

To perform a measurement on a fabric, we first perform the above protocol, then we remove the fabric from the Instron machine, lay it undisturbed on a table for ten minutes or longer, then reset it in the machine in this zero-strain state. To reproduce the measurement, we do not need to repeat the protocol. We remove the fabric from the Instron machine, lay it undisturbed on a table for ten minutes or longer, then reset it in the machine in this zero-strain state and obtain nearly identical results, see Fig.~\ref{fig:nextday}. The slight differences between the data taken on two different days can be attributed to differences in the spatial distribution of the paperclips across the needle of the clamps in the initial setup.

In Fig. \ref{fig:d*}, we note that signatures of return point memory are present in the data: in the second tests and later, a discontinuity occurs in the data once it is stretched beyond the largest extension of the previous test, for example at $14\,\textrm{mm}$ in the second test (orange).

Measurements of the same sample taken at different times are consistent. However, we  observe variations between the numerical values of the return points between \textit{different} identically prepared samples. This effect can be seen by recording the values of the return points in Fig.~\ref{fig:RPM} vs.\ Fig.~\ref{fig:rate} in the main text, which were taken on different samples that were identically prepared. We attribute these differences to imperfect control of tension during the machine knitting process and inhomogeneity along the yarn itself.

\subsection{Return Point Memory in related fabrics}\label{sec: other_fabrics}

In the main text, we present data for a ribbed sample stretched wale-wise. We show here that the phenomenology of return point memory also applies to fabrics stretched in the course direction and to fabrics of other topology.

\subsubsection{Return Point Memory in course-wise stretching direction}

\begin{figure}[b]
	\centering	\includegraphics[width=.4\textwidth]{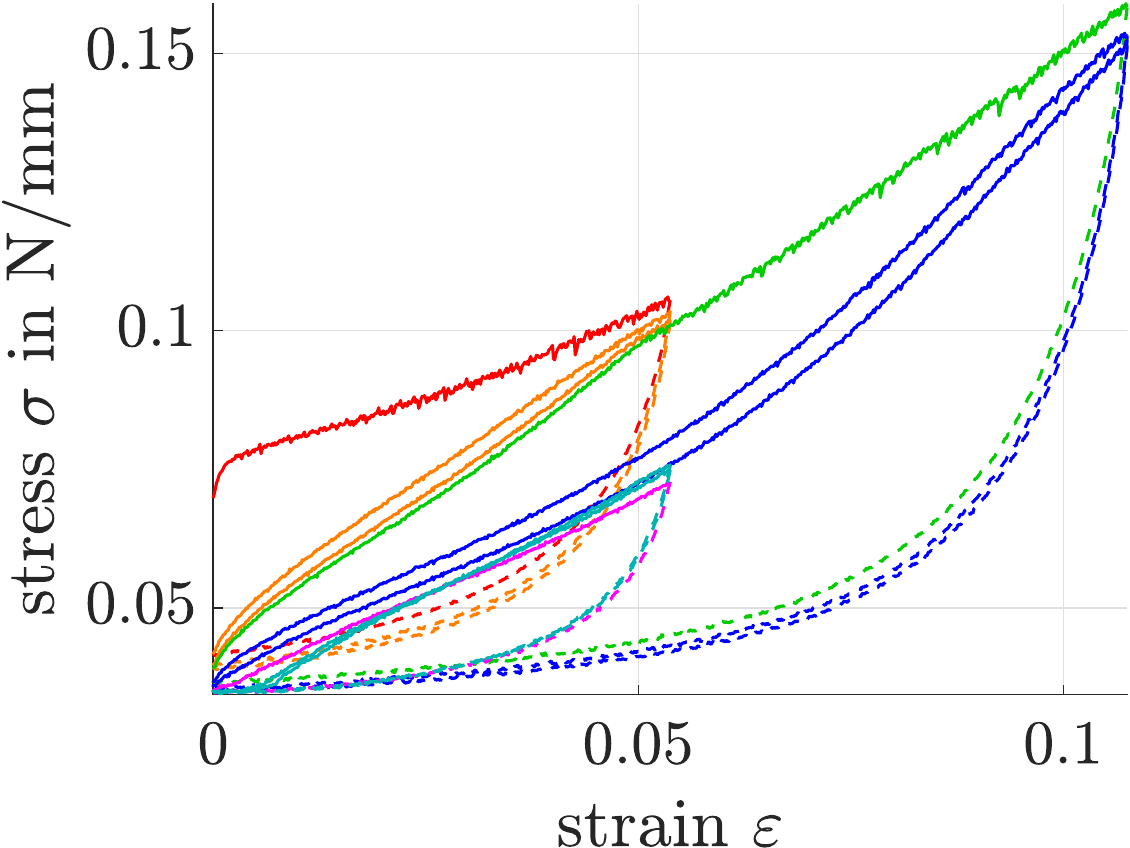}		
    \caption{Return point memory demonstration for ribbed fabric stretched in the course direction. The fabric is first stretched to $\varepsilon_1 \sim 0.055$ (red, then, orange curves), then to $\varepsilon_2 \sim 0.11$ (green, blue) and then again to $\varepsilon_1$ (pink, cyan). The slope of all loading curves is less than the slopes of corresponding curves in the wale-direction experiments, however the properties of non-local memory, wiping out, and congruency are all also present in this measurement. The measurements were performed at $\dot{\Delta} = 0.25$ mm/s for a fabric of initial length $d^* = 186$ mm.}
\label{fig:coursewise}
\end{figure}

Ribbed fabric is notoriously stretchy in the course direction, and is thus typically used on the ends of sleeves and collars of sweaters. For a ribbed sample prepared according to the fabrication protocol in \ref{sec:fabrication}, with paperclips attached in the transverse direction, we determine the zero strain state for reproducible measurements to be $d^* = 186\,\textrm{mm}$, more than twice its rest length of $d_0 = 84\,\textrm{mm}$. The demonstration of return point memory in this sample is shown in Fig. \ref{fig:coursewise}.

\begin{figure}[t]
	\centering
	\unitlength=1mm
    \begin{picture}(70,120)
        \put(0,62){\includegraphics[width =0.38\textwidth]{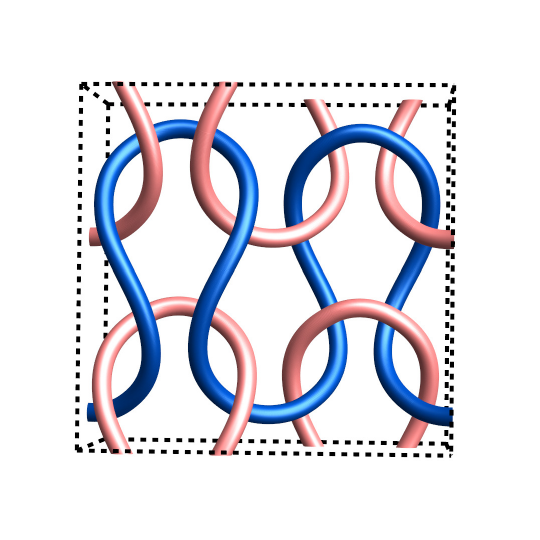}}
        \put(0,0){\includegraphics[width =0.38 \textwidth]{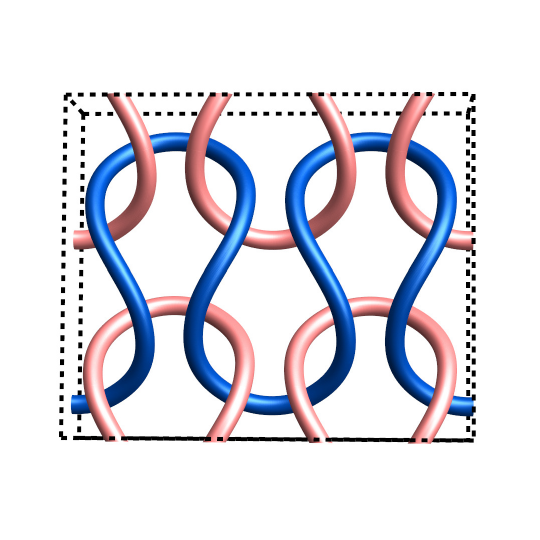}}
        \put(0,117){a)}
        \put(0,55){b)}
    \end{picture}
    \caption{{\bfseries a)}  Yarn-level schematic of ribbed fabric containing two full unit cells of the fabric stacked vertically. Each unit cell consists of one knit and one purl stitch connected horizontally. The blue and pink yarns represent yarn added to the fabric in different rows of the knitting process. {\bfseries b)}  Yarn-level schematic of stockinette fabric, consisting of four unit cells, which are one knit stitch.  Schematic images shared by Michael Dimitriyev (private communication, October 2023).}
\label{fig:rib_vs_stockinette}
\end{figure}

\subsubsection{Return Point Memory with different stitch topology: stockinette}

Our primary data is taken on ribbed samples due to their flat equilibrium shape, however we observe the same memory phenomenology for stockinette fabrics.

Ribbed and stockinette fabric differ in their topology. The stitch is the elementary unit of knitting; this is made by creating a loop of yarn and pulling it through an existing loop of the fabric. The fundamental building blocks of knitted fabrics are knit and purl stitches, where the newly created loop is pulled through the front and back of the existing loop, respectively. The unit cell of ribbed fabric is a knit and purl stitch connected in the course direction whereas the unit cell of stockinette is a single knit stitch, shown in Fig.~\ref{fig:rib_vs_stockinette}.

\begin{figure}[t]
	\centering	\includegraphics[width=.4\textwidth]{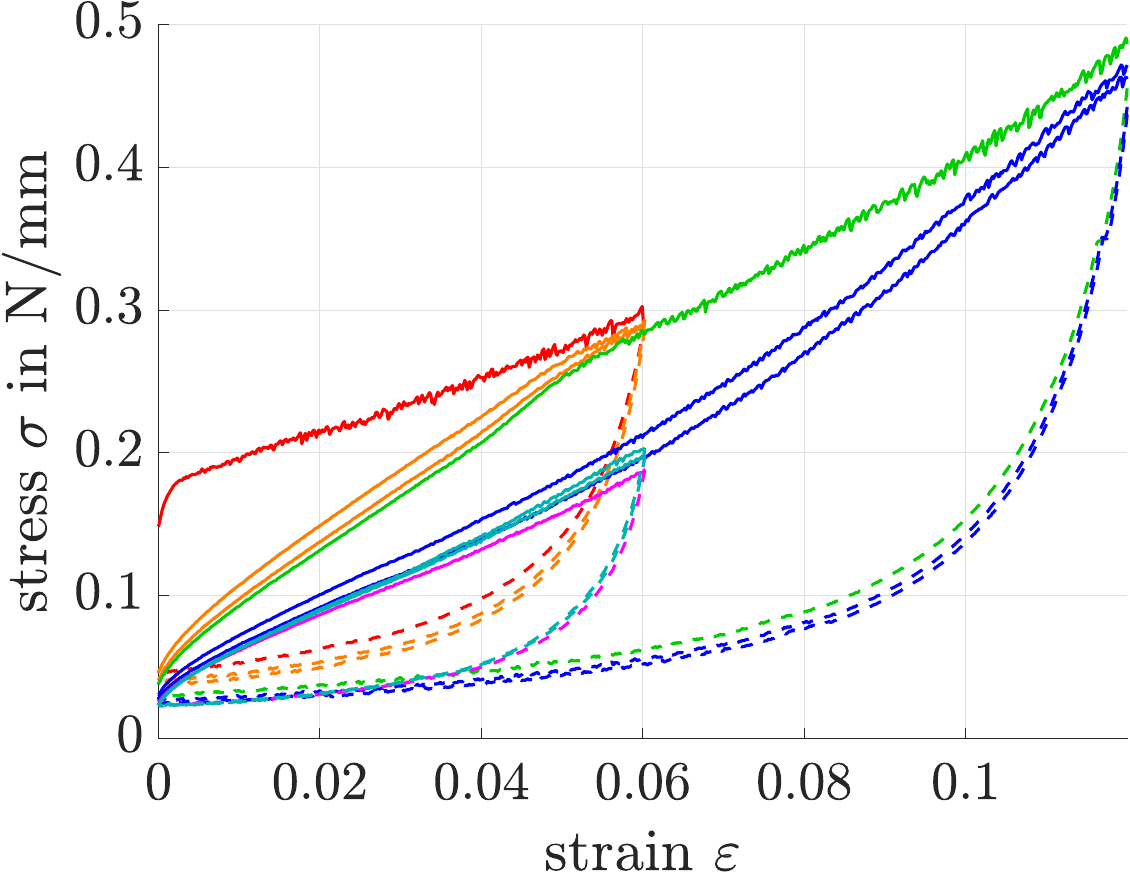}		
    \caption{Return point memory demonstration for stockinette fabric stretched in the wale direction. The fabric is first stretched to $\varepsilon_1 \sim 0.06$ (red, then, orange curves), then to $\varepsilon_2 \sim 0.12$ (green, blue) and then again to $\varepsilon_1$ (pink, cyan). The slope of all loading curves is significantly greater than the slopes of corresponding curves in the wale-direction measurements of ribbed fabric, however the properties of non-local memory, wiping out, and congruency are all also present in this measurement. The measurements were performed at $\dot{\Delta} = 0.25$ mm/s for a fabric of initial length $d^* = 166$ mm.}
    \label{fig:stockinette}
\end{figure}

While the underlying topology of these fabrics is starkly different, stockinette fabric shows the same qualitative features of return point memory as its ribbed counterpart: non-local memory, wiping-out, and congruency, as seen in Fig.\ \ref{fig:stockinette}.

\subsection{Stress-relaxation data}\label{sec:relax}

\begin{figure}[t]
	\centering	\includegraphics[width=.45\textwidth]{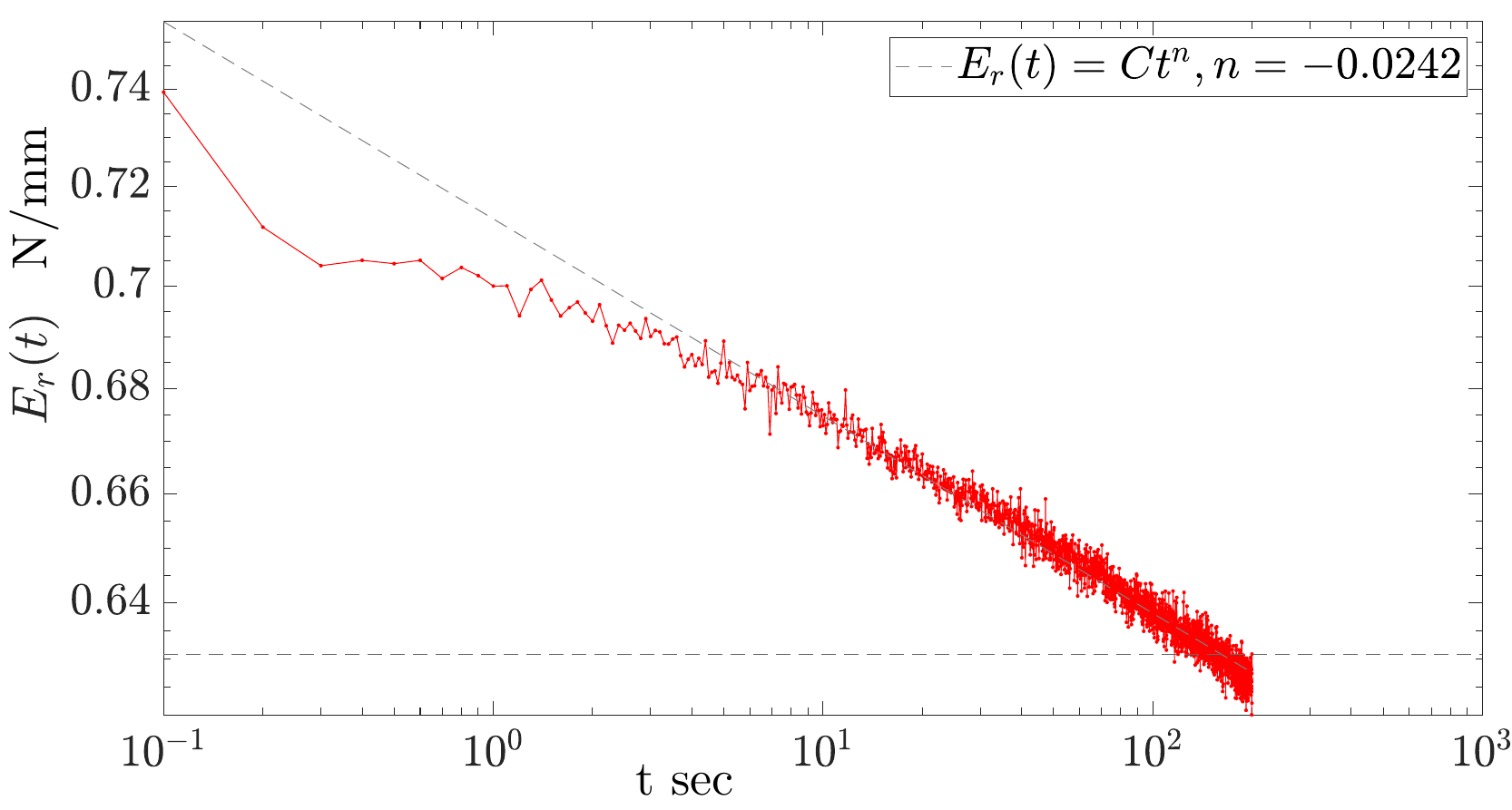}		
    \caption{Stress relaxation data for ribbed fabric stretched to $10\,\textrm{mm}$ then held at this extension for $500$ seconds. The power-law fit captures the slow relaxation of the load.}
    \label{fig:loglog}
\end{figure}

In this section, we show data for stress-relaxation of a fabric sample under a fixed strain. Specifically, as in the main text, for a ribbed fabric stretched wale-wise. In these experiments we fixed the distance between the clamps to be $204\,\textrm{mm}$, comparable to experiments described in the main text. We then strained the sample to an extension of $10\,\textrm{mm}$ at $5\,\textrm{mm/s}$ and held it for $200$ seconds. In Fig.\ \ref{fig:loglog}, we plot the stress relaxation function $E_r(t) = \sigma(t)/\varepsilon_0$ where $\varepsilon_0$ is the applied strain. We find that after a short transient time, which we estimate to be one second, the data fits well to a power-law relaxation curve $E_r(t) \sim t^{-n}$ with exponent $n = -0.0242$. Since the time scale on which stress in fabrics relaxes is long compared to our experiments, we neglect this aspect of the fabrics in our model.

\section{Supplementary Model Details}\label{sec:supmodel}
\subsection{Model details and explanation}
Proposed nearly 100 years ago by Ferenc Preisach, the Preisach model~\cite{preisach:35,mayergoyz:86,krasnoselskii.pokrovskii:89} is a fundamental achievement in the mathematical description of hysteresis. Initially developed in the context of ferromagnetism to formally relate the magnetic field and magnetization in terms of hysteresis, we adopt it here to model the experimentally observed rate-independent hysteresis in knitted fabrics. 
The Preisach model features three key aspects, which occur in knitted fabrics. First, the model shows the property of \textbf{nonlocal memory}, which is the dependency of the output (stress) not only on the current input (strain) but also on the entire history of the input. In terms of knitted fabrics, this means the same strain can lead to two different stress values depending on loading or unloading state. Second, the model shows the \textbf{wiping-out} property which means that the global extreme values of the input define the material behavior. That is, the maximal experienced stretch of the fabric dictates its behavior. Third, the \textbf{congruency} property states that all minor hysteretic loops are identical in shape and size, which coincides with the observation that once the fabric has been stretched to, say, $15\%$, then when it is stretched to $7.5\%$, the $7.5\%$-hysteresis lies inside the $15\%$-hysteresis loop.

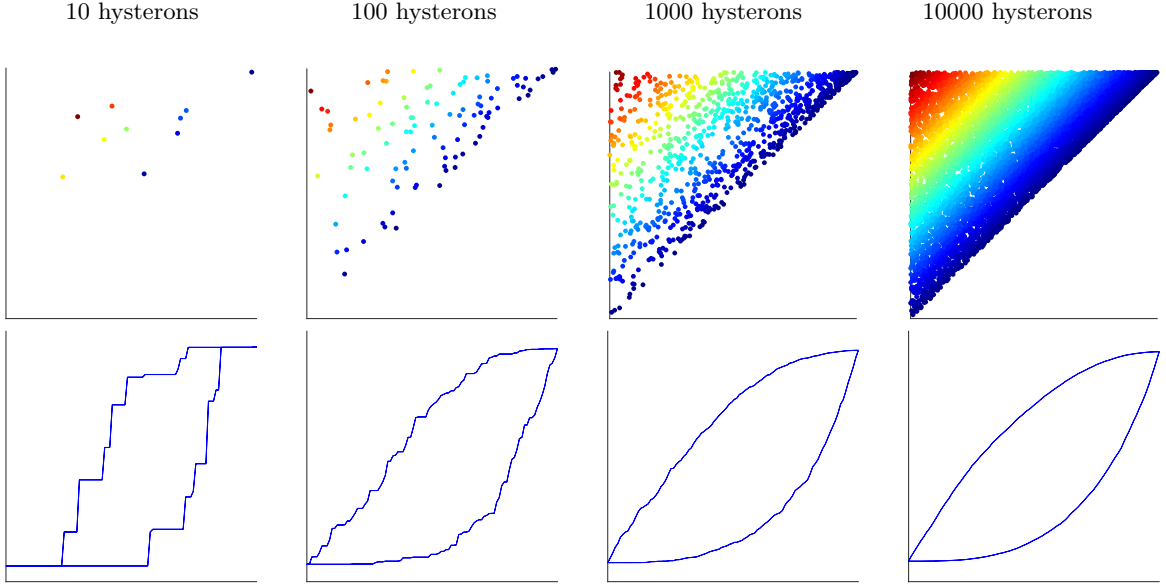
\begin{figure*}[t]
	\centering
	\unitlength=1mm
	\begin{picture}(150,80)
		\put(8,75){10 hysterons}
		\put(46,75){100 hysterons}
		\put(85,75){1000 hysterons}
		\put(122,75){10000 hysterons}
		\put(0,35){\includegraphics[width=3.35cm]{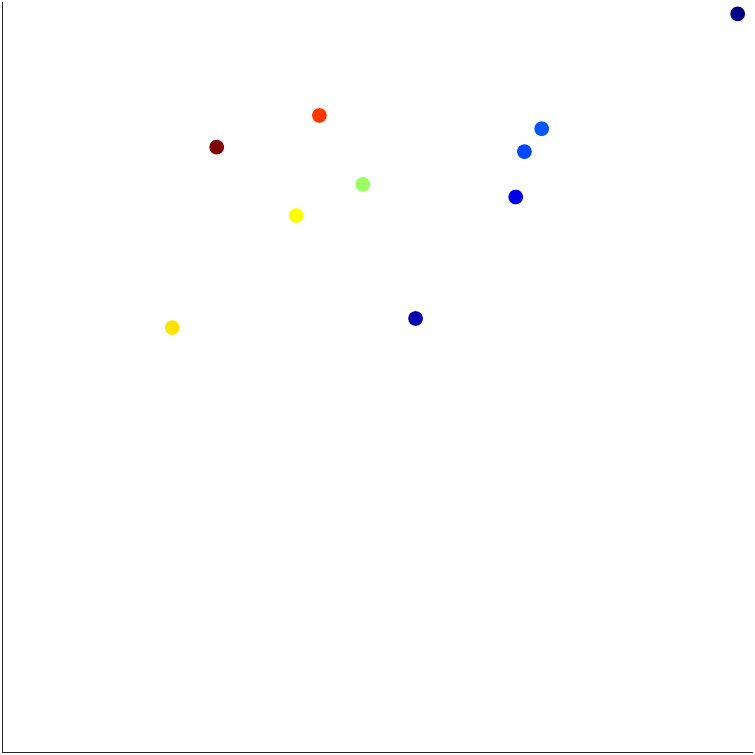}}
		\put(40,35){\includegraphics[width=3.35cm]{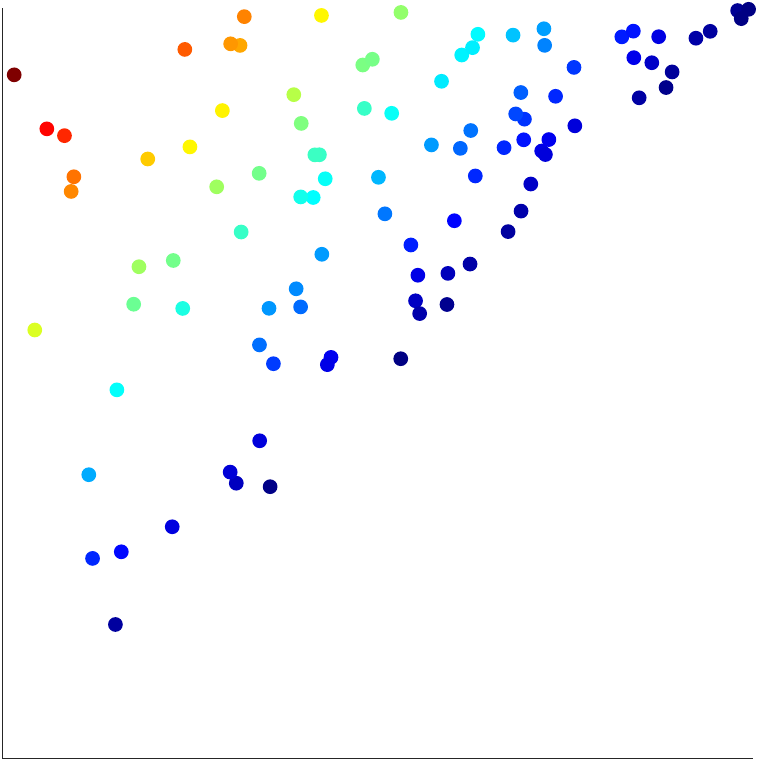}}
		\put(80,35){\includegraphics[width=3.35cm]{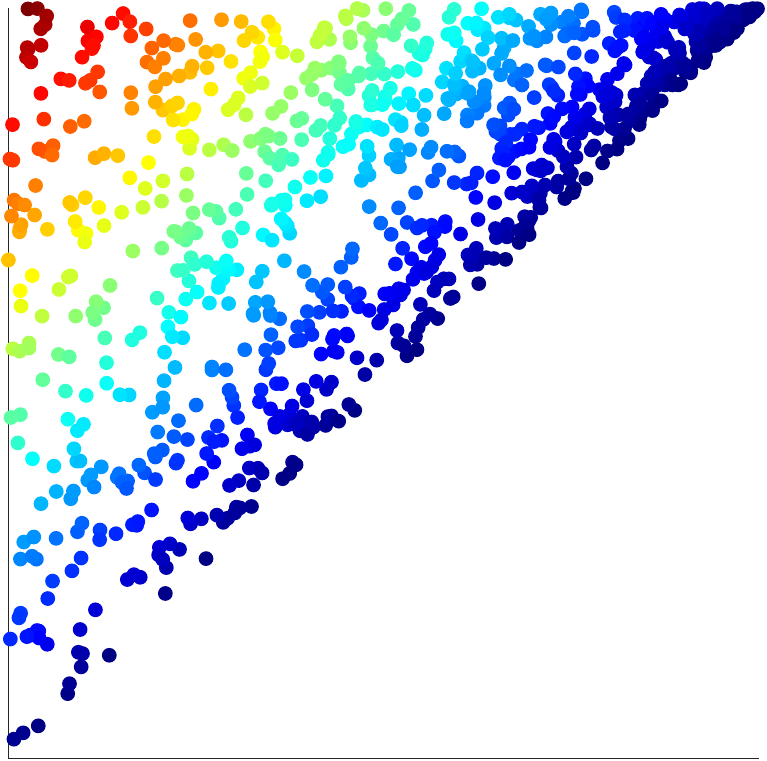}}
		\put(120,35){\includegraphics[width=3.35cm]{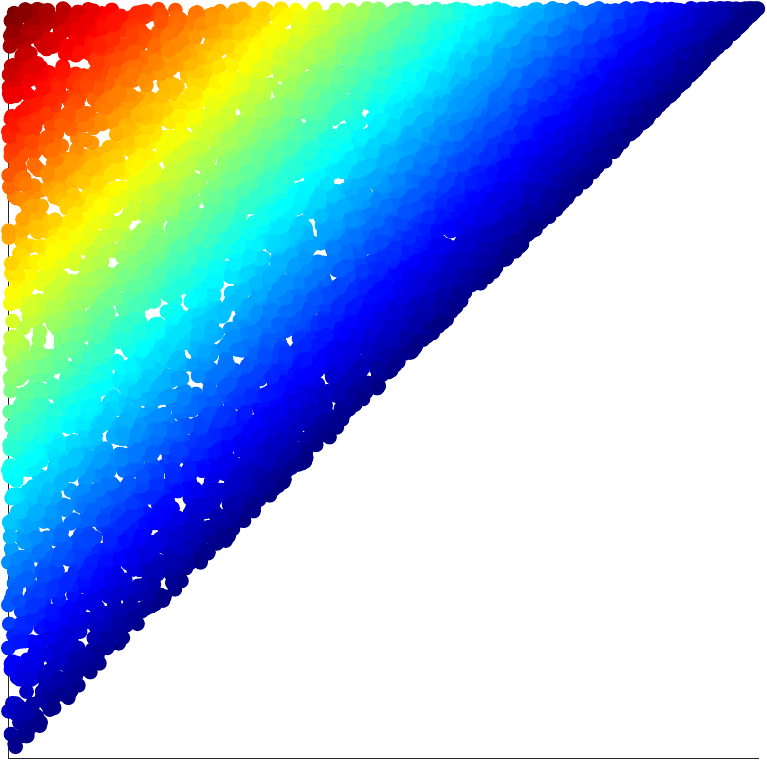}}
        \put(0,0){\includegraphics[width=3.35cm]{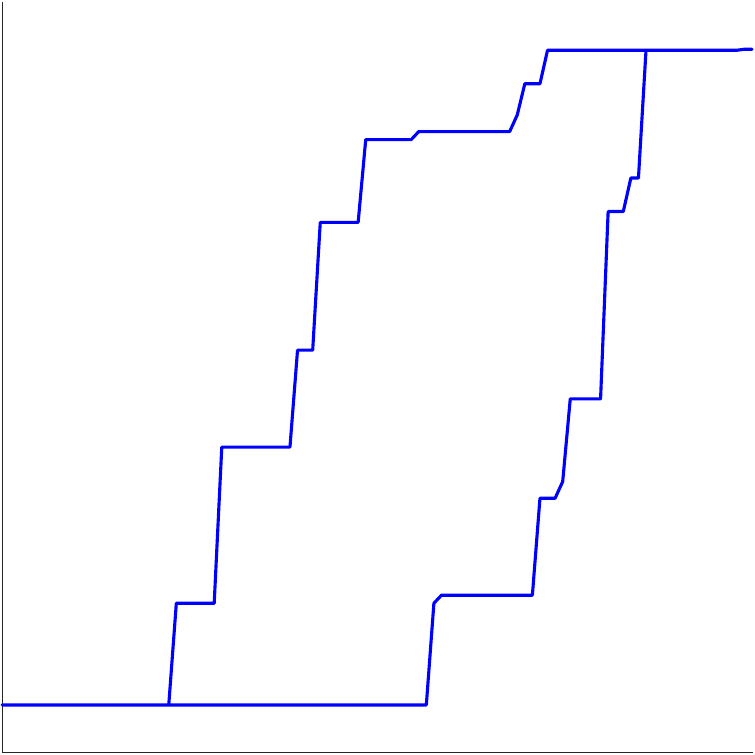}}
		\put(40,0){\includegraphics[width=3.35cm]{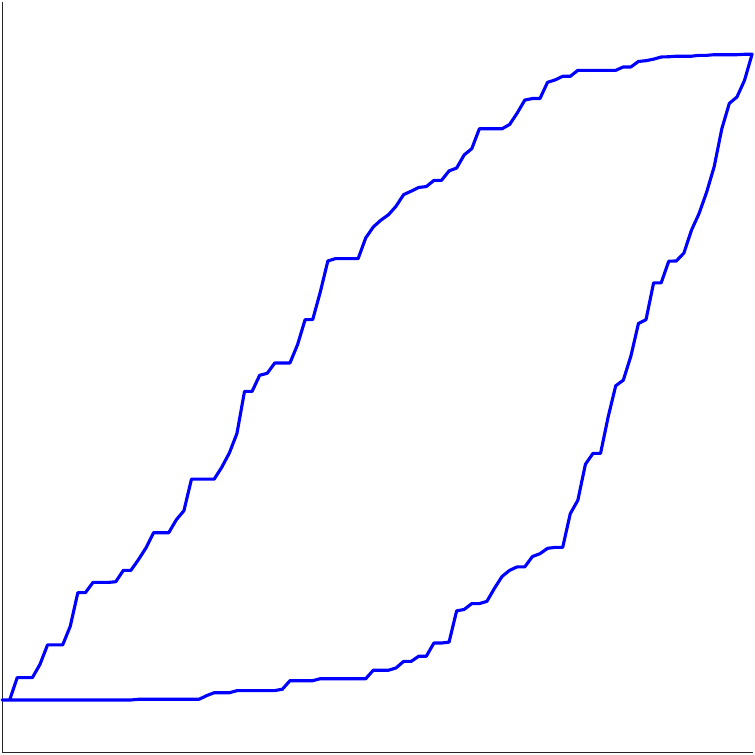}}
		\put(80,0){\includegraphics[width=3.35cm]{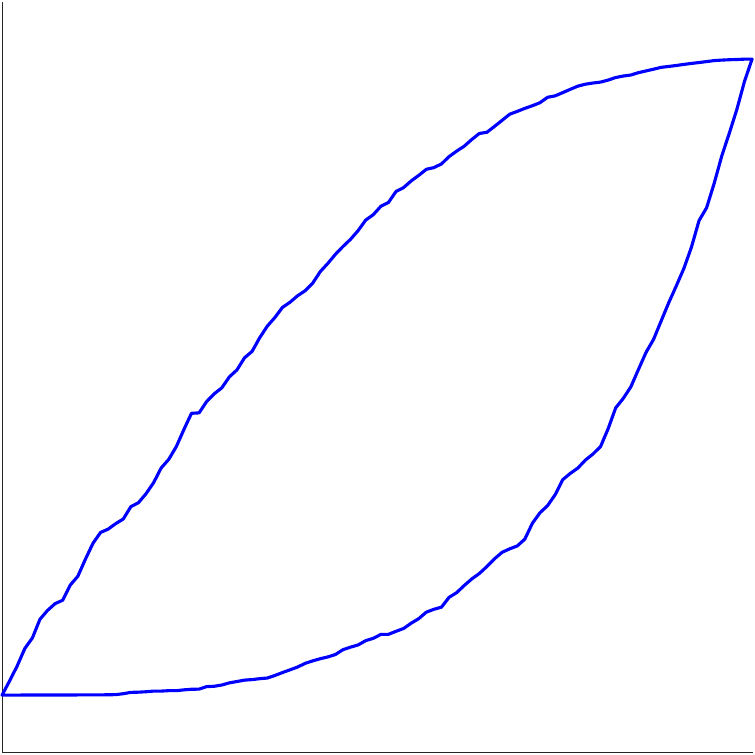}}
		\put(120,0){\includegraphics[width=3.35cm]{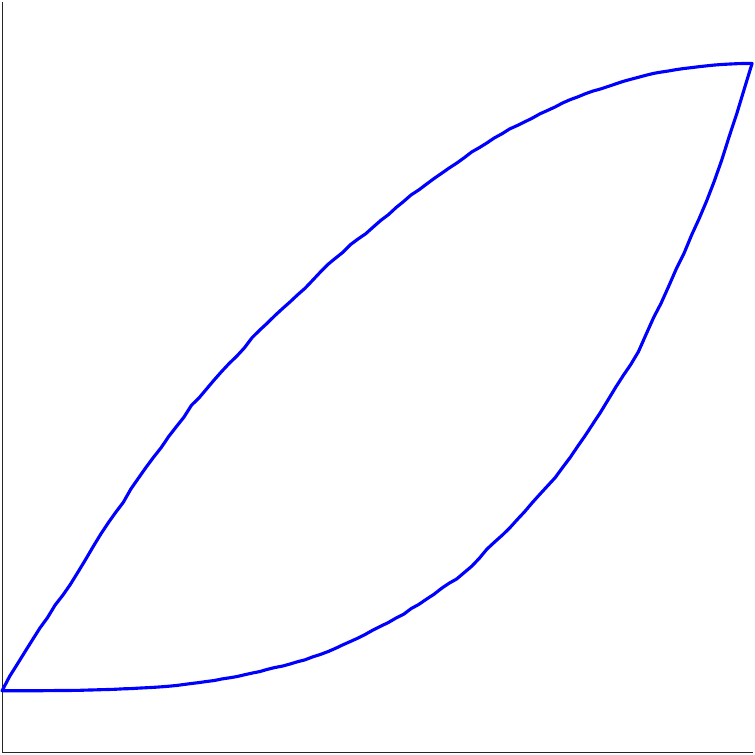}}
	\end{picture}
	\caption{Lower row: stress versus strain response corresponding to the use of a discrete number of hysterons in the $\alpha,\beta$ Preisach-plane as shown in the upper row. The more continuous the distribution, the better a stress-strain-hysteresis is approximated. Starting with a low number of hysterons, the stress-strain curve resembles a stair-case function. Considering a high number of hysterons smooths the curve.}
	\label{fig:hysterons}
\end{figure*}

Preisach suggested that such a hysteretic material response can be described by a superposition of simple rectangular-shaped relays, also known as hysterons. Each hysteron has two states, typically referred to as `on' and `off'. By assigning different threshold values for switching from `on' to `off' and vice versa, and overlaying many hysterons with varying thresholds, one can reproduce the desired hysteretic behavior. 
Explicitly, a hysteron relay or switch function in this context, having an upper~$\alpha$ and lower~$\beta$ strain threshold ($\beta\leq\alpha$), is given  
\begin{equation}
	s(\varepsilon[t]; \alpha, \beta 
    )=\begin{cases}
		+ \sigma_0 &\mathrm{if}\,\varepsilon\geq\alpha\\ -\sigma_0&\mathrm{if}\,\varepsilon\leq\beta\\
		s (\varepsilon[t^-]; \alpha, \beta)&\mathrm{else}\,,
	\end{cases}
\end{equation}
where $s(\varepsilon[t^-]; \alpha, \beta)$ represents the hysteron's stress state just prior to time $t$, $\sigma_0$ is a unit of stress, $\varepsilon$ is the input strain signal, and $s(\varepsilon[t]; \alpha, \beta)$ is the output stress state of the hysteron. 
The hysteron remains in its current state with the possibility to switch to the on-state when $\varepsilon$ passes $\alpha$ from below, or to the off-state when $\varepsilon$ passes $\beta$ from above.
Assuming a superposition of a distribution of hysterons with individual thresholds having a probability density (weights) $\mu(\alpha,\beta)$ leads to a hysteretic stress response given by
\begin{equation}
	\sigma(t) =\int\int_{\alpha\geq\beta}\mu(\alpha',\beta')s(\varepsilon[t]; \alpha', \beta')\,\mathrm{d}\alpha'\,\mathrm{d}\beta'\,,\label{dis-hys}
\end{equation}
cf. Fig.~\ref{fig:hysterons}.

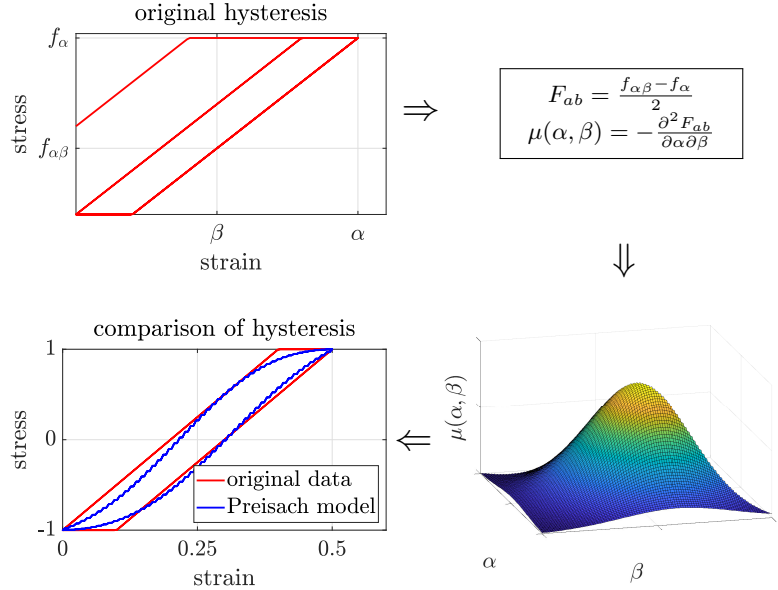
\begin{figure*}[t]
	\centering
	\unitlength=1mm
	\begin{picture}(110,75)
		\put(0,42){\includegraphics[width=5cm]{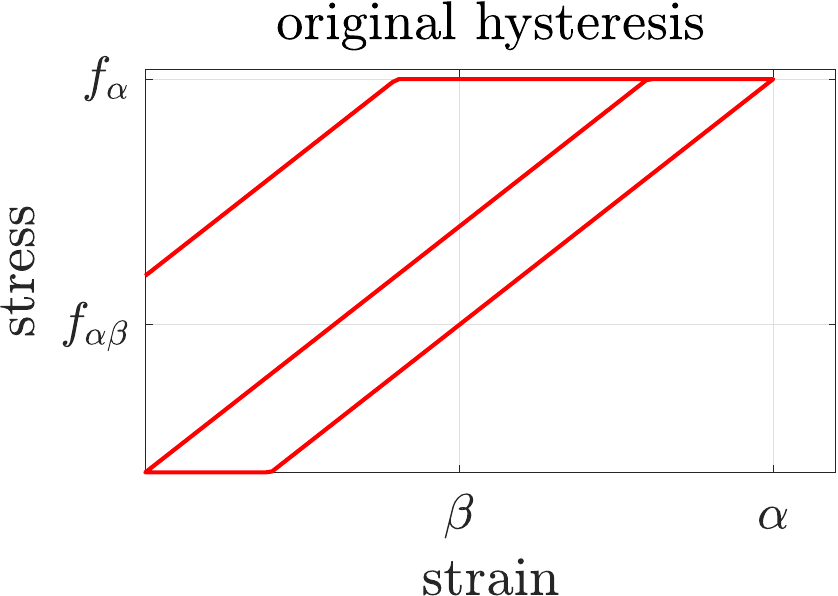}	}
		\put(0,0){\includegraphics[width=5cm]{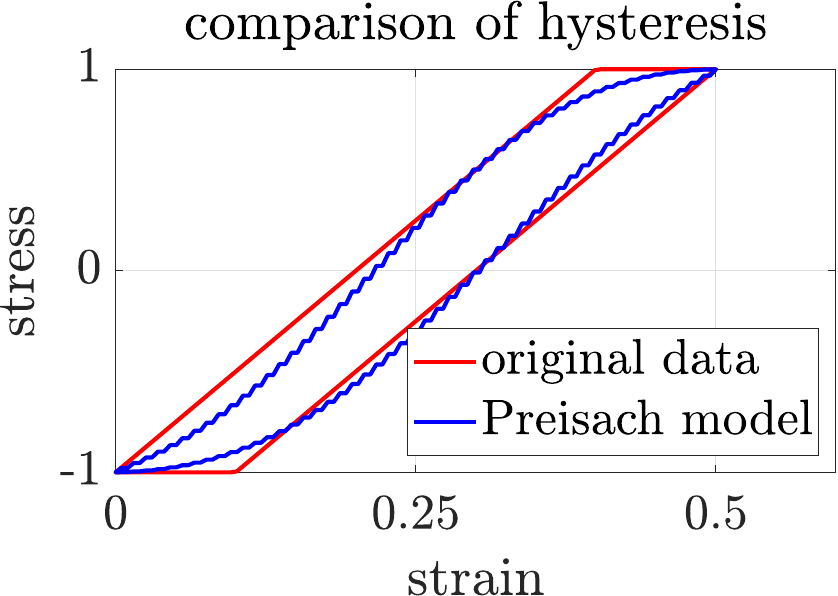}	}
		\put(56,0){\includegraphics[width=5cm]{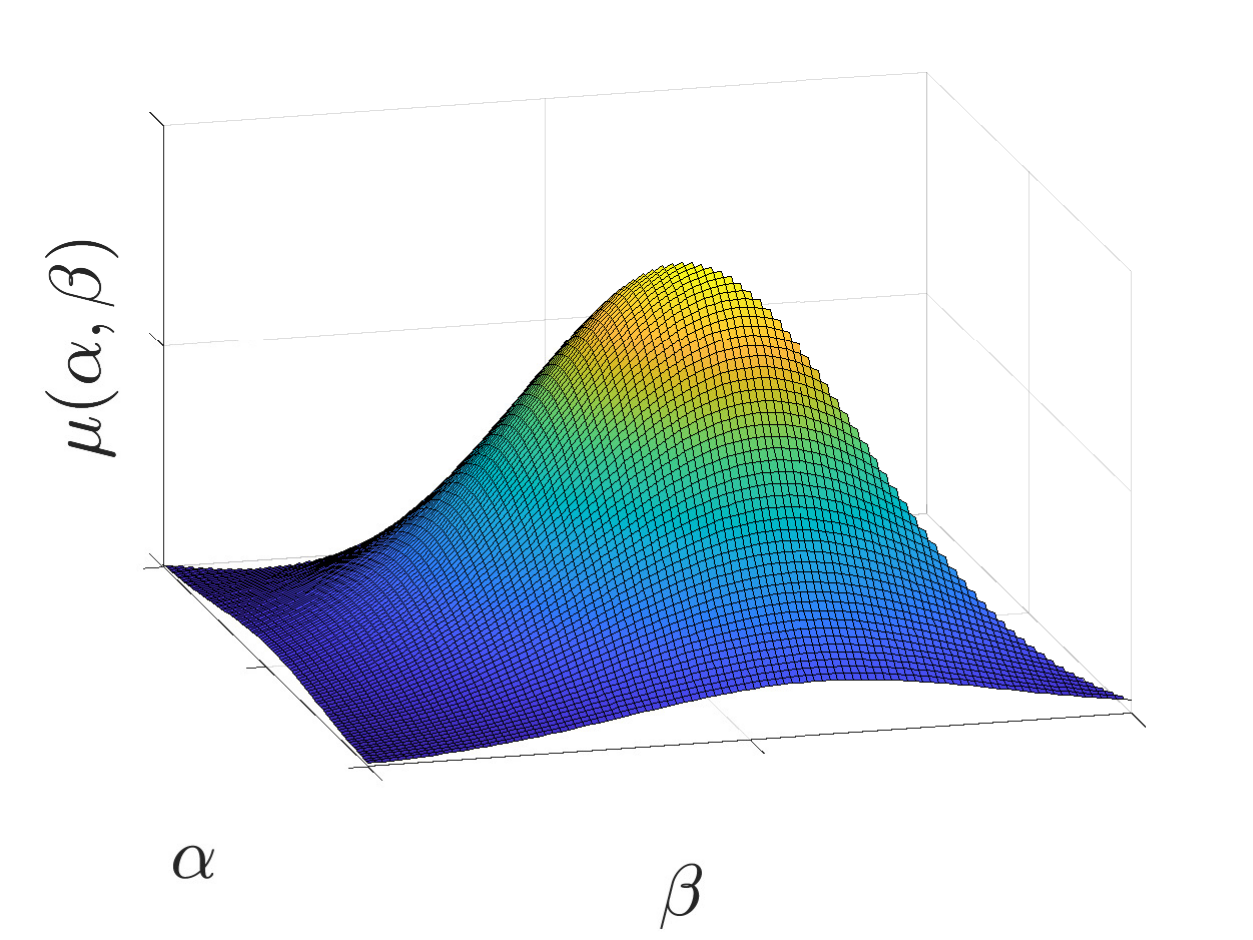}	}
		\put(52,62){\Large$\Rightarrow$}
		\put(51,18){\Large$\Leftarrow$}
		\put(80,42){\Large$\Downarrow$}				
		\put(65,62){\fbox{\parbox{3cm}{\small $ F_{ab} = \frac{f_{\alpha\beta}-f_\alpha}{2}$ $\mu(\alpha,\beta) = -\frac{\partial^2F_{ab}}{\partial\alpha\partial\beta}$}}}
	\end{picture}
    \caption{ Preisach model: Based on experimentally measured stress-strain hysteresis loops, the Preisach density function~$\mu$ is determined as second derivative of function~$F_{ab}$ with the characteristics, $\mu\geq0$ on Preisach-$\alpha$-$\beta$-plane and $\int_\alpha\int_\beta\mu(\alpha,\beta)=1$. That Preisach density function weights the hysterons in order to describe arbitrary hysteretic behavior.}
    \label{fig:preisach_a}
\end{figure*}

Fig.~\ref{fig:preisach_a} schematically sketches the idea of how one formulates mathematical relations for the Preisach model. A hysteresis loop, representing the relationship between stress and strain,  exhibits characteristic upper and lower branches for loading and unloading. If we consider a loading to an upper threshold $\alpha=a$ followed by an unloading to a lower threshold $\beta=b$, then one can define 
an auxiliary function $F_{ab}$ based on the stress at strain $a$ and the stress at strain $b$ (after loading to $a$). The mixed second derivative of this auxiliary function directly yields the Preisach density function $\mu(\alpha,\beta)$~\cite{lubarda:93}. This density function quantifies the distribution and weighting of all hysterons contributing to the overall hysteretic behavior of the material, i.e., it effectively encodes the system’s hysteretic memory.

Specifically, consider given hysteresis data and follow the increase of the input until some value $\alpha$ with output level $f_\alpha$, then upon decrease of the input to some value $\beta$ with output level $f_{\alpha\beta}$, the difference of these output values is proportional to the integral of the density function 
\begin{widetext}
\begin{equation}
    f_{\alpha\beta}-f_\alpha=2\int\int \mu(\alpha',\beta')\,\mathrm{d}\alpha'\,\mathrm{d}\beta'\quad\Rightarrow\quad\mu(\alpha,\beta)=\frac{1}{2}\frac{\partial^2 }{\partial\alpha\,\partial\beta} (f_{\alpha\beta}-f_\alpha)\,.
\end{equation}
\end{widetext}
Following this process for multiple values of $\alpha$ and $\beta$ permits the identification of $\mu(\alpha,\beta)$.

\begin{figure*}[t]
	\centering
	\unitlength=1mm
	\begin{picture}(160,65)
		\put(2,0){\includegraphics[width=0.4\textwidth]{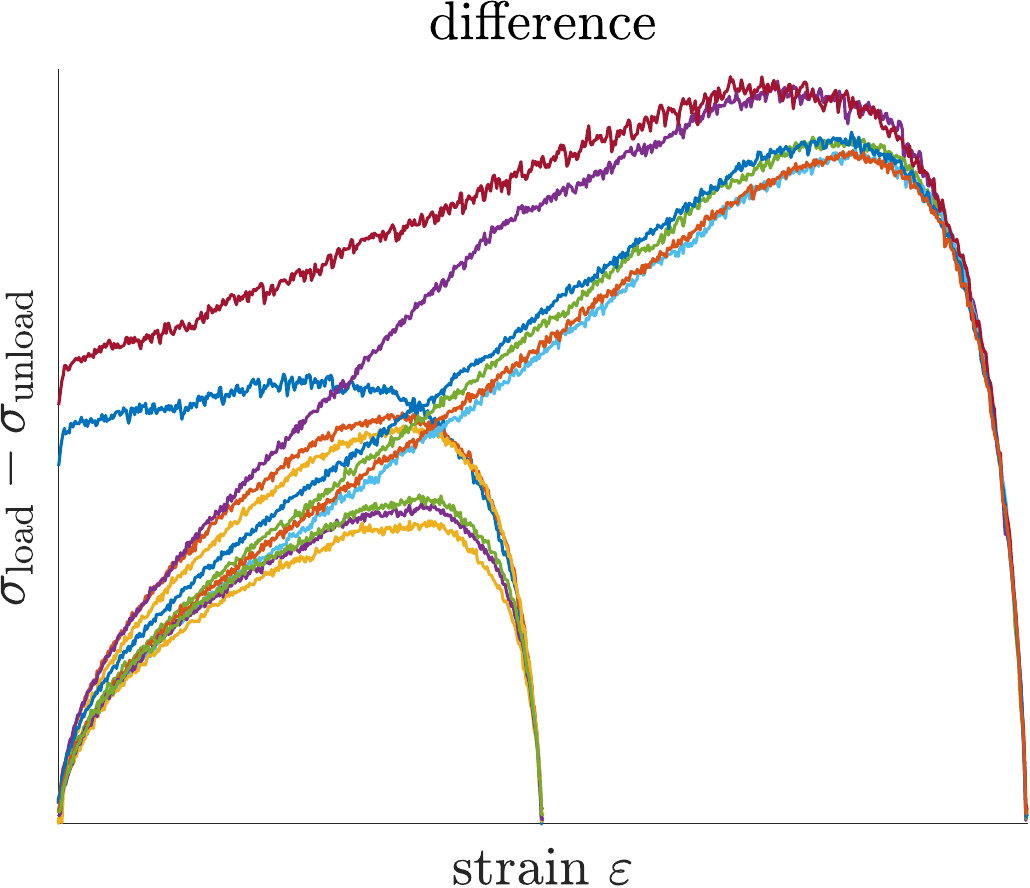}}		
		\put(87,0){\includegraphics[width=0.4\textwidth]{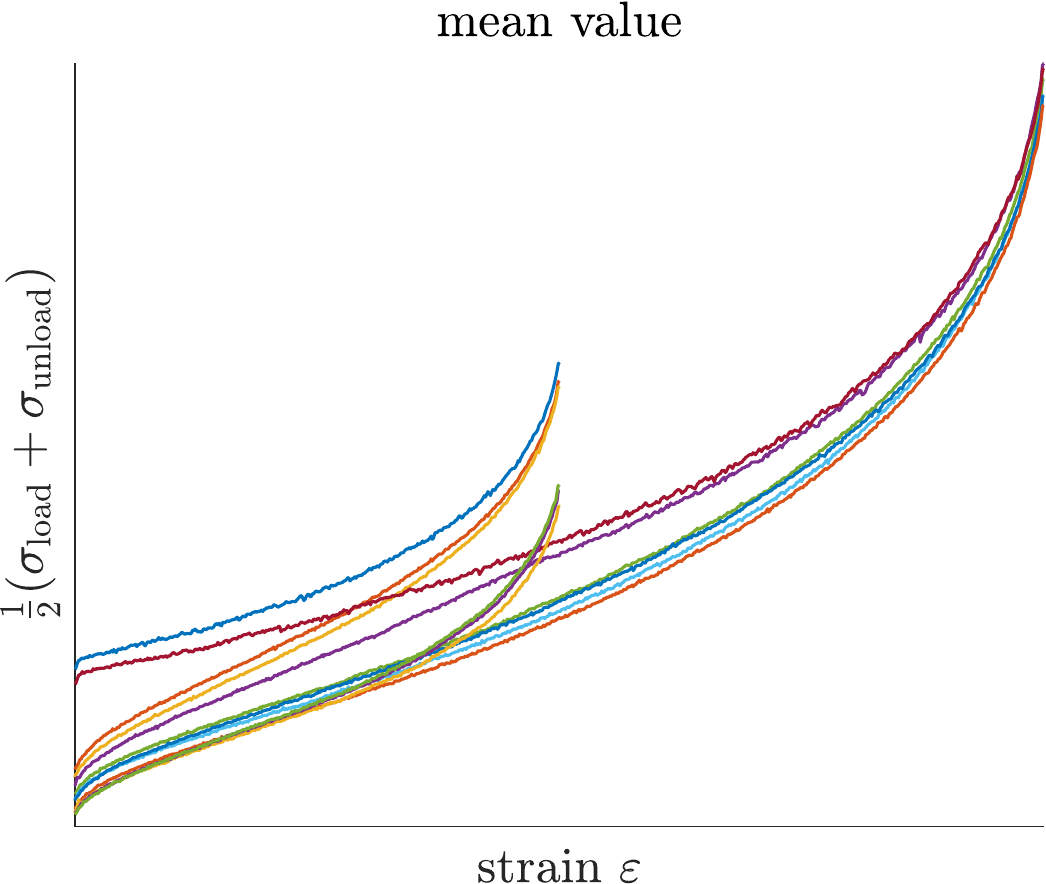}}
        \put(0,60){a)}
        \put(85,60){b)}
	\end{picture}
    \caption{Each loading-unloading cycle is analyzed individually regarding {\bfseries a)} the difference between loading and unloading stress values and {\bfseries b)} the corresponding mean value (by color). The difference of the curves is interpreted as magnitude of the rebound range~$R$, while the mean value is associated to the location of the rebound center~$R_C$.}
    \label{fig:data_analysis}
\end{figure*}

Modeling knitted fabrics is complicated by the nonlinear structure of their hysteresis and thus the basic Preisach model needs to be adapted to incorporate the observed characteristics. 
The Preisach density function $\mu(\alpha,\beta,\varepsilon_t,\varepsilon_\mathrm{max})$, for one, is non-stationary and depends on the material's history, $\varepsilon_\mathrm{max} = \max_{t'\le t} \varepsilon[t']$, and entanglement state, $\varepsilon_t$.  Secondly the switching function $s(\alpha,\beta,\varepsilon_t,\varepsilon_\mathrm{max})$ (the relay function), is dependent on these same variables.
Thus is becomes quite challenging to formulate $\mu$ and $s$ explicitly.

As a computationally simplified approximation to the Preisach model we introduce the model given in the main body of the paper, wherein the total strain is decomposed as
$\varepsilon = \varepsilon_e + \varepsilon_t$ and $\sigma = E(\varepsilon_\mathrm{max})\, \varepsilon_e$.
The evolution of the entanglement strain $\varepsilon_t$ is given with the help of a rebound inequality within a set of Karush-Kuhn-Tucker conditions: 
\begin{widetext}
\begin{equation*}
   g(\sigma,\varepsilon_t,\varepsilon_\mathrm{max})= |\sigma-R_C(\varepsilon_t,\varepsilon_\mathrm{max})|-R(\varepsilon_t,\varepsilon_\mathrm{max}) \leq 0\;; \qquad
   \dot\varepsilon_t = \gamma\, \partial g/\partial \sigma\;;\qquad \gamma g = 0\,.
\end{equation*}
\end{widetext}
The rebound range $R$ and the range center $R_C$ are inferred directly from the experimental data, just as in the basic Preisach model.
We compare the loading and unloading path of each respective cycle, see Fig.~\ref{fig:data_analysis}. The difference between these curves characterizes the rebound range~$R$, which increases and decreases proportional to the entanglement strain. 
The mean of the difference in each cycle during loading and unloading, respectively, represents the moving rebound center~$R_C$ of that range. The evolution of $R_C$ exhibits an S-shaped characteristic with increasing (entanglement) strain, so we adopt a sigmoid function for it.  The evolution of $R$ exhibits a saturation structure.  Herein, we adopt the following dependencies on the entanglement strain:
\begin{align*}
    R(\varepsilon_t,\varepsilon_\mathrm{max}) &= R_0 + R^*\,(1-\exp(-\delta\,\varepsilon_t))\,+\,h(\varepsilon_\mathrm{max})\,\varepsilon_t\,,\\
    R_C(\varepsilon_t,\varepsilon_\mathrm{max}) &= R_{C0}(\varepsilon_\mathrm{max})\,+ \,R_C^*\, \frac{{\varepsilon_t}^m}{{\varepsilon_t}^m+\varepsilon_0^m(\varepsilon_\mathrm{max})}\,.
\end{align*}
The dependency on the maximum strain history is expressed by 
\begin{align*}
	E&=E_0(1+E_f\,\varepsilon_\mathrm{max})\,,
	&&h = h_0\frac{1}{1+h_f\,\varepsilon_\mathrm{max}}\,,\\
	R_{C0}&=R_{C0}^A(1-\exp(R_{C0}^B\,\varepsilon_\mathrm{max}))\,,
	&&\varepsilon_0=\varepsilon_0^A+\varepsilon_0^B\,\varepsilon_\mathrm{max}\,.
\end{align*}

The material parameters given in Table~\ref{tab:parameter} are found in a fitting to the experimental data from the test in Fig.~\ref{fig:RPM}b. The \texttt{fmincon} function (i.e., function minimization with constraints) in Matlab is used with options listed in Table~\ref{tab:fmincon} to find the minimum of 
\begin{equation}
    \min_v f(v) = \frac{1}{N_\mathrm{data}}\sum_{i=1}^{N_\mathrm{data}}(f_{\mathrm{sim}_i}-f_{\mathrm{exp}_i})^2 
\end{equation}
with $v$ as the material parameters in Table~\ref{tab:parameter} using suitable bounds, $N_\mathrm{data}$ as the number of data points, and $f_{\mathrm{sim}_i}$ and $f_{\mathrm{exp}_i}$ as simulated and experimental results, respectively.

To ensure the positiveness of the dissipation during computation, we check every cycle and verify that the dissipated energy is positive 
\begin{equation}
	W_\mathrm{diss} = \oint \sigma\,\mathrm{d}\varepsilon_t\qquad\Rightarrow\qquad\Delta W_\mathrm{diss} = \sigma\,\Delta\varepsilon_t\,.
\end{equation}
\begin{table}[h]
\caption{Fitted material parameters.}
\label{tab:parameter}
\begin{tabular}{|p{4.6cm}|c|p{1.3cm}|c|}
        \hline
			initial Young's modulus &$E_0$&$3.8$&MPa\\
			adaptivity of $E$ &$E_f$&$2.31$&MPa\\\hline
			initial rebound range &$R_0$&$0.00125$&MPa\\
			limit of rebound range&$R^*$&$0.003$&MPa\\
			linear material parameter &$h_0$&$0.35$&-\\
			adaptivity of $h$ &$h_f$&$2.2$&-\\
            exponential material parameter &$\delta $&$200$&-\\\hline
			rebound center &$R_{C0}^A$&$-0.027$&MPa\\
			&$\phantom{x}R_{C0}^B\phantom{x}$&$-17$&MPa\\
			&$R_C^*$&$21$&MPa\\
			&$m$&$1.45$&-\\\hline
			sigmoid function&$\varepsilon_0^A$&$4$&-\\
			&$\varepsilon_0^B$&$18.7$&-\\\hline
	\end{tabular}
\end{table}

\begin{table}[h]
\caption{Parameters for \texttt{fmincon} for fitting data in Tab.~\ref{tab:parameter}.}
\label{tab:fmincon}%
\begin{tabular}{|p{2cm}|p{1.8cm}|p{1.8cm}|p{1.8cm}|}
    \hline
	MaxFunction\linebreak Evaluations & Max\linebreak Iterations & Optimality\linebreak Tolerance & Step\linebreak Tolerance\\\toprule
    1e5 & 1e3 & 1e-8 & 1e-12\\\hline
	\end{tabular}

\end{table}

\end{document}